# Coding Theorem and Converse for Abstract Channels with Time Structure and Memory

Martin Mittelbach and Eduard A. Jorswieck, *Senior Member, IEEE*

*Abstract*—A coding theorem and converse are proved for a large class of abstract stationary channels with time structure including the result by Kadota and Wyner (1972) on continuous-time real-valued channels as special cases. As main contribution the coding theorem is proved for a significantly weaker condition on the channel output memory – called total ergodicity w. r. t. finite alphabet block-memoryless input sources – and under a crucial relaxation of the measurability requirement for the channel. These improvements are achieved by introducing a suitable characterization of information rate capacity. It is shown that the $\psi$-mixing output memory condition used by Kadota and Wyner is quite restrictive and excludes important channel models, in particular for the class of Gaussian channels. In fact, it is proved that for Gaussian (e. g., fading or additive noise) channels the $\psi$-mixing condition is equivalent to finite output memory. Further, it is demonstrated that the measurability requirement of Kadota and Wyner is not satisfied for relevant continuous-time channel models such as linear filters, whereas the condition used in this paper is satisfied for these models. Moreover, a weak converse is derived for all stationary channels with time structure. Intersymbol interference as well as input constraints are taken into account in a general and flexible way, including amplitude and average power constraints as special case. Formulated in rigorous mathematical terms complete, explicit, and transparent proofs are presented. As a side product a gap in the proof of Kadota and Wyner – illustrated by a counterexample – is closed by providing a corrected proof of a lemma on the monotonicity of some sequence of normalized mutual information quantities. An operational perspective is taken and an abstract framework is established, which allows to treat discrete- and continuous-time channels with (possibly infinite input and output) memory and arbitrary alphabets simultaneously in a unified way.

*Index Terms*—Channels with memory, coding capacity, coding theorem, continuous-time channels, discrete-time channels, general alphabets, infinite memory, mixing conditions, Shannon theory, weak converse

## I. INTRODUCTION

*Motivation.* Since Shannon started information theory a large number of publications has been devoted to the mathematically rigorous derivation of channel coding theorems. The goal was to develop the classical Shannon theory for communication models of increasing complexity and generality capturing more and more practically relevant situations and phenomena. The complexity of such models is determined, for instance, by the space of symbols allowed for communication and the time structure of the signals, i. e., finite vs. infinite alphabets and discrete vs. continuous time. Furthermore, complexity is increased by taking technical constraints, e. g., for the transmit signals, into account or by incorporating aspects such as intersymbol interference or memory effects of the random noise process, which disturbs the transmission.

A main motivation for this paper is to establish an abstract and unifying framework, that allows us to formulate a general coding theorem for a point-to-point communication link in a mathematically rigorous way under practically useful assumptions. A central objective is to include general continuous-time continuous-valued transmission models because much less attention is payed to this case compared to discrete models. Moreover, the goal is a reduction to the essential channel properties required to prove the coding statements, particularly including conditions for infinite memory. In the literature different regularity conditions are considered when coding results are generalized, where some of the conditions are difficult to verify in practical situations, however, after detailed analysis it turns out they are too restrictive, e. g., concerning the measurability of the channel transition probabilities. As can be read in Ahlswede's lecture notes [2, p. 323] "It therefore seems that a thorough investigation 'of the regularity conditions' needed for extending Shannon's coding theory to the continuous case should not only be of interest for the mathematician in his ivory tower, but also to the communication engineer".

*Main contributions.* The main contribution of this paper is a generalization of a coding theorem and a converse of Kadota and Wyner [3] with regard to channel model, input constraints, definition of information rate capacity, and, most notably, w. r. t. required channel properties. Kadota and Wyner considered a continuous-time channel with real-valued input and output signals, whereas we consider channels with time structure in general, i. e., simultaneously discrete- as well as continuous-time channels with completely arbitrary alphabets (Definition 2.4). Further restrictions of the alphabets, e. g., to separable metric spaces as in [4], are not required to derive the results. To prove the coding theorem (Theorem 3.1) we require analog to [3] the considered channel to be stationary, causal, and asymptotically input-memoryless (Definition 2.7). However, with respect to output memory we achieve a significant generalization. Kadota and Wyner used a property they called asymptotic output-memorylessness, which is introduced later as $\psi$-mixing condition (Definition 4.3). By introducing an alternative characterization of information rate capacity (Definition 2.19) we are able to prove the coding theorem under a much weaker condition called total ergodicity w. r. t. finite alphabet block-memoryless input sources (Definition 2.7). Due to the employed definition of information capacity we also have the advantage, that we can use a relaxed measurability condition called (M) for the channel (Paragraph 2.6), which we can easily verify for important continuous-time channel models such as linear filters (Example 3.4). In contrast, the measurability condition (M′) (Section III.B) required by Kadota and Wyner excludes these relevant models, even if the filter response has finite support, i.e., if the channel has finite input memory. Furthermore, we prove that the measurability considered in this paper is implied by the channel input memory condition required anyway to prove the coding theorem (Lemma 2.8 and Paragraph A.1). A corresponding result does not exist for the condition (M′) used by Kadota and Wyner.

To explicitly demonstrate that the $\psi$-mixing output memory condition of Kadota and Wyner is quite restrictive – in particular for the important class of Gaussian channels – we prove that for Gaussian channels the $\psi$-mixing condition is equivalent to finite output memory (Theorem 4.7 and Paragraph B.2). As a result and in contrast to the coding theorem in this paper, Kadota and Wyners's formulation of the coding theorem is for example not applicable to the important stationary additive Gaussian noise channel with proper rational noise spectral density (Example 4.8).

We prove a weak converse for all stationary channels with time structure (Theorem 3.2) based on a convenient representation of the information rate capacity (Lemma 2.21 and Paragraph A.3). No further restrictions on the channel properties or alphabets are required. Due to a generic characterization we have a convenient flexibility in taking intersymbol interference into account. Furthermore, input constraints are incorporated in an abstract form. In contrast to [3], we are able to accomplish this without using a standard extension of the channel input $\sigma$-algebra. Additional requirements for the input constraints such as compactness like in [4] are not required. Only some regularity over time called (R) is needed (Paragraph 2.5). Outer measures as in [5] can also be avoided due to the relaxed measurability condition, both making derivations simpler and more transparent. We also close a gap in the proof in [3] regarding a lemma on the monotonicity of some sequence of normalized mutual information quantities (Lemma 2.16). The lemma is directly used in the proof of the coding theorem (Theorem 3.1) and indirectly (via Lemma 2.21) in the proof of the weak converse (Theorem 3.2). We provide a corrected proof (Paragraph A.2) and show by an explicit counterexample (Appendix C) that the argumentation in [3] is incorrect. The results are obtained in a general framework by a suitable and consistent combination of various classical tools such as Feinstein's lemma or Pinsker's ergodic theorem of information theory. The combination of the coding theorem and the weak converse establishes the information rate capacity (Definition 2.19) as the coding capacity for a large class of channels (Paragraph 2.13 and Remark 3.3). A more detailed discussion on the explicit contributions of the paper is given in Section V after precisely presenting and deriving the results.

*More related work.* Classical works on coding theorems and converse theorems for discrete-time finite alphabet channels with memory are due to Khinchin [6, Part II], Takano [7], Feinstein [8], Wolfowitz [9], Adler [10], Pfaffelhuber [11], and Gray and Ornstein [12]. An extension of Khinchin's results [6] to discrete-time channels with continuous alphabets and finite capacity was obtained by Rosenblatt-Roth [13], [14]. Jacobs [15] further developed the results of Rosenblatt-Roth [13] and added a weak converse. A coding theorem and a weak converse for discrete-time channels with abstract alphabets were also derived by Wagner [16], who considered the stationary memoryless case. In contrast to all previously mentioned references Wagner took input constraints into account and used Gallager's error exponent in his proof of the coding theorem, whereas the other authors used the maximal coding method, i.e., Feinstein's lemma. Very general results for discrete-time memoryless channels with abstract alphabets, including a strong converse, were derived by Augustin [17] and Kemperman [18]. Dobrushin [19], [20] developed a generalized coding theorem for discrete-time channels from the point of view proposed by Kolmogorov [21]. He showed that information stability of the channel is sufficient for the validity of the coding theorem. Ding [22], [23] proved that this theoretical condition is also necessary when the capacity expression of Dobrushin is employed. Since this channel information stability condition is often very difficult to verify for concrete models, coding theorems under various memory properties were derived.

A well-known reference on information measures with generalized alphabets is the book of Gray [24], however, the material is restricted to standard alphabets, which is inadequate to study general continuous-time channel models. The final result on coding for discrete-time channels [24, Th. 14.1] is a revised version of the coding theorem derived in [12], weakening stationarity to asymptotic mean stationarity. Both theorems are derived under a condition called $\bar{d}$-continuity, which is less restrictive than the asymptotic input-memorylessness required in the coding theorem of this paper. However, the $\bar{d}$-continuity condition is only applicable to discrete-time finite-alphabet channels and Gray's theorems are proved for this special case only. Furthermore, [24, Th. 14.1] is a "one-shot" theorem, which assumes that the channel is used only once, whereas in this paper the repeated transmission of codewords (taking intersymbol interference into account) is considered. Kieffer [25] introduced the class of stationary weakly continuous channels. He showed that it includes all stationary $\bar{d}$-continuous channels and still allows to prove a coding theorem. However, in the sense of a joint source/channel coding theorem, whereas we focus on a pure channel coding theorem in this paper.

A general formula for the coding capacity of a discrete-time channel, including nonstationary and information unstable channels with abstract alphabets, is given by Verdú and Han [26]. The classical information-theoretic work on continuous-time transmission models is devoted to the additive Gaussian noise channel. The main objective was to give rigorous derivations of and operational significance to Shannon's celebrated capacity formula "$W \log(1 + P/NW)$", heuristically derived in [27, Part IV]. Contributions in this direction are due to Ash [28], [29], Yoshihara [30], Wyner [31], [32], Holsinger [33], Gallager [34, Ch. 8], Cordaro and Wagner [35], Baker [36], [37], [38], [39], McKeague [40], and Baker and Ihara [41]. Ihara [42], [43] extended the results by taking feedback into account. The theorems in this paper apply to a general class of continuous-time channel models whereas the previously listed literature is restricted to additive noise or additive Gaussian noise channels. Note that many authors use the method proposed by Holsinger [33] and Gallager [34, Ch. 8], where a continuous-time channel is represented by an infinite series of parallel discrete-time channels. Using a consequent measure-theoretic description and following the approach in [3] we are able to avoid this transformation completely. Therefore, we can treat discrete- and continuous-time channels with abstract alphabets in a unified way, which we believe is an important argument in favor of the path taken in this paper.

In more recent work on continuous-time channels Charalambous [44] is concerned with the control of continuous-time linear Gaussian systems over wireless fading channels with additive white noise. Weissman [45] studied mismatched estimation in continuous-time additive white Gaussian noise channels and characterized in [46] the capacity of a class of input memoryless continuous-time channels with feedback via directed information. The identification capacity of continuous-time Poisson and Gaussian channels is analyzed by Burnashev [47]. Durisi et al. [48] derived bounds on the noncoherent information capacity of continuous-time underspread wide-sense stationary uncorrelated scattering channels that are selective both in time and frequency. For those channels Durisi et al. [49] characterized the sensitivity of the capacity w. r. t. to the channel model. Furthermore, Jung [50] established a capacity formula for continuous-time doubly-dispersive channels with additive Gaussian noise. Koch and Lapidoth [51, p. 5946] and Durisi et al. [49, p. 6376] raised the question whether a coding theorem does exist for the information capacity expressions they are studying. The paper at hand provides a general framework to analyze this problem for the discrete- as well as the continuous-time fading channel considered there. How this can be done is demonstrated in [5, Par. 17.6], where for some multipath fading channels with infinite input memory, i. e., with an infinite number of path, all channel properties required in the coding theorem of this paper, have been verified. Therewith we have positive examples in the case of [51, p. 5946]. The derivations in [5, Par. 17.6] are also useful as guideline to investigate the continuous-time channels in [49] in this regard.

## II. Model and Preliminaries

In this section, we provide the material required to formulate and prove a coding theorem and a weak converse for abstract channels with time structure. We introduce a general stochastic transmission and communication model including a corresponding coding capacity as well as model properties relevant to derive the main theorems. To fix notation and also to reveal the general scope of the paper, we restate the necessary general definitions of information measures for random variables with abstract alphabet. Supporting material on information rates is provided and proved and an adequate form of information rate capacity is introduced. Required information-theoretic tools such as Feinstein's lemma and the ergodic theorem of information theory are restated in tailored form. We begin the section by establishing some general notation used throughout the paper. Basic notions of probability and measure theory are used freely. As a compact reference in this regard please refer to [52, Chs. 1–4, 6], where the notation is close to that introduced below. A compact collection of more advanced and frequently used background material from probability and measure theory (e. g., on Markov kernels and chains, densities of measures, Fubini's theorem, characterizations of ergodicity, etc.) is provided in [5, Appendix A and B].

### II.A. General Notation

*Sets, $\sigma$-algebras, (product) measurable spaces and functions.* As usual $\mathbb{Z}$, $\mathbb{N}$, $\mathbb{N}_0$, and $\mathbb{R}$ denote the set of integers, positive integers, nonnegative integers, and real numbers, respectively. Throughout this paper we denote by $T$ the set of time indices. Whenever the index set $T$ is used it can be replaced either by $\mathbb{Z}$ to model discrete-time or by $\mathbb{R}$ to model continuous-time. Occasionally, it is convenient to extend $T$ and consider $\overline{T} = T \cup \{-\infty, \infty\}$. The sets of positive and nonnegative time indices are denoted by $T_+$ and $T_0$, respectively. In case of $T = \mathbb{Z}$ we use the interval notation for subsets of $T$ to denote the set of consecutive integers contained in the interval, e. g., if $v \in T_+$, then $(0, v]$ is the short hand version of $\{1, 2, \ldots, v\}$. We write $\varnothing$ for the empty set. For a set $A$ we denote by $A^c$ its complement and by $2^A$ its power set, i. e., the set of all subsets of $A$.

A pair $(\Omega, \mathcal{F})$ consisting of a space $\Omega$ and a $\sigma$-algebra $\mathcal{F}$ on $\Omega$ is called measurable space. A partition of $(\Omega, \mathcal{F})$ is any countable family $\{A_1, A_2, \ldots\}$ of disjoint sets $A_i \in \mathcal{F}$ whose union is equal to $\Omega$. If $\mathcal{G}$ is a family of subsets of $\Omega$, then $\sigma(\mathcal{G})$ denotes the smallest $\sigma$-algebra containing $\mathcal{G}$. Suppose $(\Omega, \mathcal{F})$ and $(X, \mathcal{X})$ are measurable spaces and $f$ is a function on $\Omega$ with values in $X$. If $f$ is measurable w. r. t. the $\sigma$-algebras $\mathcal{F}$ and $\mathcal{X}$, i. e., $\{f^{-1}(A) : A \in \mathcal{X}\} \subset \mathcal{F}$, then we say $f$ is $\mathcal{F}/\mathcal{X}$-measurable. In the special case where $f$ is a real-valued or numerical function, i. e., $X = \mathbb{R}$ or $X = \mathbb{R} \cup \{-\infty, \infty\}$ and $\mathcal{X}$ is the corresponding Borel-$\sigma$-algebra, then we simply say $f$ is $\mathcal{F}$-measurable.

Assume that $(\Omega_1, \mathcal{F}_1)$ and $(\Omega_2, \mathcal{F}_2)$ are measurable spaces. Then $(\Omega_1 \times \Omega_2, \mathcal{F}_1 \otimes \mathcal{F}_2)$ denotes the corresponding product measurable space consisting of the product space $\Omega_1 \times \Omega_2$ and the product $\sigma$-algebra $\mathcal{F}_1 \otimes \mathcal{F}_2$.

*Probability and measure spaces, random variables.* The triple $(\Omega, \mathcal{F}, \mu)$ consisting of the space $\Omega$, the $\sigma$-algebra $\mathcal{F}$, and the measure $\mu$ on $(\Omega, \mathcal{F})$ is called measure space. If $\mu$ is a probability measure, then $(\Omega, \mathcal{F}, \mu)$ is called probability space. Suppose $\nu$ is another measure on $(\Omega, \mathcal{F})$. Then we write $\mu \ll \nu$ if $\mu$ is absolutely continuous w. r. t. $\nu$, i. e., if every $\nu$-nullset is a $\mu$-nullset. By $\delta_\omega$ we denote the Dirac measure on $(\Omega, \mathcal{F})$ concentrated at some $\omega \in \Omega$. If $(\Omega_1, \mathcal{F}_1, \mu_1)$ and $(\Omega_2, \mathcal{F}_2, \mu_2)$ are two measure spaces, then $\mu_1 \otimes \mu_2$ denotes the product measure obtained from $\mu_1$ and $\mu_2$, which is defined on the product measurable space $(\Omega_1 \times \Omega_2, \mathcal{F}_1 \otimes \mathcal{F}_2)$

Let $(\Omega, \mathcal{F}, P)$ be a probability space. For a random variable $\xi$ defined on $(\Omega, \mathcal{F}, P)$ we denote by $P_\xi$ the distribution of $\xi$. If $\xi$ has values in the measurable space $(X, \mathcal{X})$, then $X$ is called alphabet of the random variable $\xi$. If $\{\xi_t, t \in T\}$ is a family of random variables, where $\xi_t$ has values in $(X_t, \mathcal{X}_t)$ and $(X_t, \mathcal{X}_t) = (X_0, \mathcal{X}_0)$ for all $t \in T$, then $X_0$ is also called alphabet of the family. For any set $F \in \mathcal{F}$ and $\sigma$-algebra $\mathcal{A} \subset \mathcal{F}$ we denote by $\mathrm{P}(F|\mathcal{A})$ the conditional probability of $F$ given $\mathcal{A}$. If $\eta$ is a another random variable on $(\Omega, \mathcal{F}, \mathrm{P})$ with values in $Y$ and $A \in \mathcal{X}$, then $\mathrm{P}_{\xi|\eta}(A|\cdot)$ denotes the conditional probability of $\{\xi \in A\}$ given $\eta$, which is a nonnegative random variable on $(Y, \mathcal{Y}, \mathrm{P}_\eta)$. If $\xi$ and $\eta$ and a third random variable $\zeta$ on $(\Omega, \mathcal{F}, \mathrm{P})$ form a Markov chain in this order, then we write $(\xi - \eta - \zeta)$. If $\xi$ and $\eta$ are real or numerical random variables, then $\mathrm{cor}(\xi, \eta)$ is the (normalized) correlation of $\xi$ and $\eta$.

*Logarithms.* If we write $\log$, then we always assume logarithms w.r.t. base $e$. In addition, we always suppose $0 \log \frac{0}{x} = 0$ for $x \geq 0$ and $x \log \frac{x}{0} = +\infty$ for $x > 0$.

### II.B. CHANNEL MODEL AND CHANNEL PROPERTIES, TIME STRUCTURE

Starting with a completely general stochastic transmission model – the channel – we introduce time structure by considering Cartesian products as channel input and output spaces. We introduce an abstract model to specify constraints on the channel input signals, define channel properties directly relevant in connection with proving a coding theorem and converse, and give useful related implications.

**(2.1) Definition (Channel).** Let $X$ be a set and $(Y, \mathcal{Y})$ be a measurable space, both arbitrary. Assume that $\kappa$ is a function on $X \times \mathcal{Y}$ with values in $[0, 1]$ such that $\kappa(x, \cdot)$ is a probability measure on $(Y, \mathcal{Y})$ for all $x \in X$. Then $\kappa$ is called a channel with input space $X$ and output measurable space $(Y, \mathcal{Y})$.

**(2.2) Remark (Interpretation, measurability, input-output measure).** The interpretation of the definition of a channel is as follows: The quantity $\kappa(x, B)$ specifies the probability that the received symbol lies in the set $B \in \mathcal{Y}$ given the transmitted symbol was $x \in X$.

In the literature, the definition of an abstract channel typically includes a $\sigma$-algebra $\mathcal{X}$ on the input space $X$ and it is required that $\kappa(\cdot, B)$ is an $\mathcal{X}$-measurable function for all $B \in \mathcal{Y}$. With this additional technical condition $\kappa$ is a Markov kernel from $(X, \mathcal{X})$ to $(Y, \mathcal{Y})$, which is the mathematical model used, e.g., in [53, Sec. 1.5][1], [55, Sec. 6.2], [3], [56, Sec. 3.1], or [24, Sec. 2.2] and also in [5]. For the model of a discrete-time memoryless channel Ahlswede pointed out in [2, Sec. 6.1] that the measurability condition can be omitted when coding problems are studied, as was already demonstrated in [17, p. 11]. Also [18] and [57] considered the discrete-time memoryless case without requiring measurability.

Subsequently, we will see that for the general coding result of this paper on abstract channels with discrete or continuous time structure and (infinite) memory some form of measurability of the channel is required. However, in contrast to the assumptions in the literature, the relaxed condition considered in this paper (Paragraph 2.6) is satisfied (and easily verified) for relevant continuous-time channel models such as linear filters (Example 3.4). Further, it is automatically satisfied for the channel properties needed anyway to prove the coding theorem (Lemma 2.8). The measurability aspect of the channel is discussed in detail in Section III.B.

In the derivations below we often consider a subset $A \subset X$ of the channel input space together with a $\sigma$-algebra $\mathcal{A}$ on $A$ such that $\kappa(\cdot, B)$ as function on $A$ is $\mathcal{A}$-measurable for all $B \in \mathcal{Y}$. Then $\kappa$ is a Markov kernel from $(A, \mathcal{A})$ to $(Y, \mathcal{Y})$ and we can induce with a probability measure $\mu$ on $(A, \mathcal{A})$ the probability measure $\mu\kappa$ on the product space $(A \times Y, \mathcal{A} \otimes \mathcal{Y})$

$$\mu\kappa(C) = \int_A \kappa(x, C_x) \, \mathrm{d}\mu(x), \qquad C \in \mathcal{A} \otimes \mathcal{Y}, \qquad (1)$$

where $C_x = \{y \in Y : (x, y) \in C\} \in \mathcal{Y}$ denotes the $x$-section of the set $C$. We call $\mu$ (channel) input probability measure and $\mu\kappa$ (channel) input-output probability measure. The marginal measure of $\mu\kappa$ on $(Y, \mathcal{Y})$ given by $\nu(B) = \mu\kappa(A \times B)$ for all $B \in \mathcal{Y}$ is called (channel) output probability measure.

Note, for $\kappa$ to be a Markov kernel from $(A, \mathcal{A})$ to $(Y, \mathcal{Y})$ it is sufficient that $\kappa(\cdot, G)$ as a function on $A$ is $\mathcal{A}$-measurable for all $G \in \mathcal{G}$, where $\mathcal{G}$ is some set system generating $\mathcal{Y}$, i.e., $\mathcal{Y} = \sigma(\mathcal{G})$, which is closed w.r.t. finite intersections.

**(2.3) Time structure, inverse images, shifts.** Let $\{X_t, t \in T\}$ be a family of sets with $X_t = X_0$ for all $t \in T$. For $u \leq v \in \overline{T}$ we denote the product space generated by the subfamily related to the index set $(u, v]$ by[2]

$$X_u^v = \underset{t \in (u,v]}{\times} X_t. \qquad (2)$$

With $x_u^v = \{x_t, t \in (u, v]\}$, where $x_t \in X_t$ we denote an element of $X_u^v$. As short hand notation we use $X$, $X_-^v$, and $X_u^+$ for $X_{-\infty}^\infty$, $X_{-\infty}^v$, and $X_u^\infty$, respectively. The same convention applies to $x_u^v$.

For the inverse image of a set $A \subset X_u^v$ w.r.t. the projection from $X$ onto $X_u^v$ we write

$$[A] = \{x \in X : x_u^v \in A\}.$$

We naturally extend this notation to a family $\mathcal{A}$ of subsets of $X_u^v$ on an element-by-element basis, i.e.,

$$[\mathcal{A}] = \{[A] : A \in \mathcal{A}\}.$$

The $w$-shifted version of an element $x_u^v = \{x_t, t \in (u, v]\} \in X_u^v$ for some $w \in T$, denoted by $\langle x_u^v \rangle_w$, is an element of $X_{u+w}^{v+w}$ given by

$$\langle x_u^v \rangle_w = \{\tilde{x}_s, s \in (u+w, v+w]\}, \qquad \tilde{x}_s = x_{s-w}.$$

We naturally extend this notation on an element-by-element basis to a set $A \subset X_u^v$ and a family $\mathcal{A}$ of subsets of $X_u^v$ to obtain the $w$-shifted versions $\langle A \rangle_w \subset X_{u+w}^{v+w}$ and $\langle \mathcal{A} \rangle_w$. Additionally, if $\mathcal{A}$ is a $\sigma$-algebra on $A$, then the $w$-shifted copy $\langle \mu \rangle_w$ of a measure $\mu$ on $(A, \mathcal{A})$ is a measure on $(\langle A \rangle_w, \langle \mathcal{A} \rangle_w)$ defined by

$$\langle \mu \rangle_w(F) = \mu(\langle F \rangle_{-w})$$

---

[1] Russian original, see [20, Sec. 1.5] for English and [54, Sec. 1.5] for German translation.

[2] Considered as subset of $T$ the index set $(u, v]$ denotes an half-open interval if $u < v < \infty$, the open interval $(u, \infty)$ if $u < v = \infty$, the set $\{v\}$ if $u = v \neq \pm\infty$, and the empty set $\varnothing$ otherwise.

for any $F \in \langle \mathcal{A} \rangle_w$. Using the shift operator $\langle \cdot \rangle_w$ allows us to emphasize the position on the time axis $T$. A set $A \subset X$ is called $w$-invariant if

$$A = \langle A \rangle_w = \{\langle x \rangle_w : x \in A\}$$

holds and (shift-)invariant, if it is $w$-invariant for all $w \in T$.

Further, we consider the family $\{(Y_t, \mathcal{Y}_t), t \in T\}$ of measurable spaces with $(Y_t, \mathcal{Y}_t) = (Y_0, \mathcal{Y}_0)$ for all $t \in T$ and the corresponding product space and product $\sigma$-algebra

$$Y_u^v = \underset{t \in (u,v]}{\times} Y_t, \qquad \mathcal{Y}_u^v = \underset{t \in (u,v]}{\bigotimes} \mathcal{Y}_t, \qquad (3)$$

generated by the subfamilies related to the index set[2] $(u, v]$. The notation introduced above for product spaces applies in an analog manner.

Let $\nu$ be a probability measure on the product measurable space $(Y, \mathcal{Y})$ and assume that $s \in T$. Then $\nu$ is called $s$-stationary if

$$\nu(B) = \nu(\langle B \rangle_s)$$

holds for all $B \in \mathcal{Y}$ and it is called $s$-ergodic if

$$\nu(B) = 0 \quad \text{or} \quad \nu(B) = 1 \qquad (4)$$

holds for all $s$-invariant sets $B \in \mathcal{Y}$. The probability measure $\nu$ is called (strictly) stationary if it is $s$-stationary for all $s \in T$. Further, $\nu$ is called ergodic if (4) holds for all invariant sets $B \in \mathcal{Y}$.

For any $t \in T$ let $\eta_t$ be a random variable on $(\Omega, \mathcal{F}, P)$ with values in $(Y_t, \mathcal{Y}_t)$. Then $\eta_u^v$ denotes the random variable on $(\Omega, \mathcal{F}, P)$ with values in $(Y_u^v, \mathcal{Y}_u^v)$, which is defined for any $\omega \in \Omega$ by

$$\eta_u^v(\omega) = \{\eta_t(\omega), t \in (u, v]\}.$$

We identify $\eta_u^v$ with the family $\{\eta_t, t \in (u, v]\}$ of random variables, which is (a segment of) a random process either with discrete time (i.e., a random vector or sequence) or continuous time. As short hand notation we also use $\eta$, $\eta_-^v$, and $\eta_u^+$ instead of $\eta_{-\infty}^\infty$, $\eta_{-\infty}^v$, and $\eta_u^\infty$.

The process $\eta = \{\eta_t, t \in T\}$ is called $s$-stationary (stationary, $s$-ergodic, ergodic), if the distribution $P_\eta$ of $\eta$ is $s$-stationary (stationary, $s$-ergodic, ergodic).

**(2.4) Definition (Channel with time structure).** A channel $\kappa$ with abstract input product space $X$ and output product measurable space $(Y, \mathcal{Y})$ as introduced in Paragraph 2.3 is called a channel with time structure. If we have $T = \mathbb{Z}$ for the time index set, then $\kappa$ is called discrete-time channel and if $T = \mathbb{R}$, then $\kappa$ is called continuous-time channel. Elements of $X$ and $Y$ or of corresponding sub-product spaces are called input and output (time) signals, respectively. The spaces $X_0$ and $Y_0$ are called input and output alphabet, respectively.

In the rest of the paper, $\kappa$ is a channel with time structure with input product space $X$ and output product measurable space $(Y, \mathcal{Y})$. The notation introduced in Paragraph 2.3 is used freely. It is used in an analog manner for other product spaces.

**(2.5) Input constraints, regularity condition (R).** Usually signals used for information transmission have to meet certain constraints that result for example from technical specifications. For a transmission of duration $s \in T_+$ starting at time 0, we model such constraints in an abstract way by a set $E_s \subset X_0^s$, i.e., a subset of all possible input signals in the time period $(0, s]$. From $E_s$ we build the set

$$E_s^* = \underset{k \in \mathbb{Z}}{\times} \langle E_s \rangle_{ks} \subset X \qquad (5)$$

of all signals in $X$ that satisfy the input constraint in any time period $(ks, (k+1)s]$, $k \in \mathbb{Z}$. A signal from $E_s^*$ represents a sequence of transmissions, each of duration $s$ and satisfying the input constraint. In cases of practical importance the set $E_s$ contains the null-signal such that we can represent a transmission of finite duration also by a signal from $E_s^*$. The union of all shifted versions of $E_s^*$ is the shift-invariant set

$$E_s' = \underset{t \in (0,s]}{\bigcup} \langle E_s^* \rangle_t. \qquad (6)$$

Assume now that we have a constraint set for any time duration, i.e., suppose we have the family $\mathcal{E} = \{E_s \subset X_0^s, s \in T_+\}$ of sets. Subsequently, we use the shift-invariant set

$$E' = \underset{s \in T_+}{\bigcup} E_s' \qquad (7)$$

with $E_s'$ as in (6). We are especially interested in families $\mathcal{E}$ of constraint sets, that satisfy the regularity condition (R)

$$\underset{k=0}{\overset{n-1}{\times}} \langle E_s \rangle_{ks} \subset E_{ns} \qquad (8)$$

for any $s \in T_+$ and $n \in \mathbb{N}$. This condition is important, for example, in the coding theorem (and converse) for channels with time structure. A useful sufficient condition for (8) is that

$$E_u \times \langle E_v \rangle_u \subset E_{u+v} \qquad (9)$$

holds for all $u, v \in T_+$.

As an example consider for fixed $s \in T_+$ a constraint set $E_s \subset X_0^s$ of the form[3]

$$E_s = \{\Phi_s \leq 1\} = \{x \in X_0^s : \Phi_s(x) \leq 1\}, \qquad (10)$$

where $\Phi_s$ is a nonnegative functional on $X_0^s$, also called cost function. First, let $\Phi_s$ be given by

$$\Phi_s(x) = \underset{t \in (0,s]}{\sup} \phi(x_t), \quad x = \{x_t, t \in (0,s]\} \in X_0^s, \qquad (11)$$

where $\phi$ is a nonnegative function on $X_0$ (resp. on $X_t$ for all $t \in T$). If we have real-valued input signals, i.e., $X_0 = \mathbb{R}$, and $\phi(x_t) = |x_t|$, then (11) represents the important amplitude constraint. Alternatively, let the functional $\Phi_s$ be given by

$$\Phi_s(x) = \begin{cases} \dfrac{1}{s} \displaystyle\int_{(0,s]} \phi(x_t) \, d\lambda(t) \\ c \quad \text{for } T = \mathbb{R} \text{ if } \{\phi(x_t), t \in (0,s]\} \\ \quad \text{is not Lebesgue-measurable} \end{cases}, \qquad (12)$$

for all $x = \{x_t, t \in (0, s]\} \in X_0^s$, where $\phi$ is a nonnegative function on $X_0$ (resp. on $X_t$ for all $t \in T$), c is an arbitrary constant larger than 1, and $\lambda$ denotes the counting measure on the integers if $T = \mathbb{Z}$ and the one-dimensional Lebesgue measure if $T = \mathbb{R}$. For real-valued input signals and $\phi(x_t) = x_t^2$, the functional in (12) represents the important average energy or power constraint.

---

[3] Without loss of generality we can choose 1 as upper bound in (10) because we can normalize the inequality.

It can easily be shown that the family $\mathcal{E} = \{E_s \subset X_0^s, s \in T_+\}$ of constraint sets, where $E_s$ is defined by (10) and (11) or by (10) and (12) satisfies the regularity condition (R). In the latter case we can further show that the set

$$F = \Big\{ x = \{x_t, t \in T\} \in X :$$
$$\limsup_{s \to \infty} \frac{1}{2s} \int_{(-s,s]} \phi(x_t) \, \mathrm{d}\lambda(t) \leq 1 \Big\}$$

is shift-invariant and that

$$E' \subset F$$

holds, with $E'$ as given in (7). For example, if the input signals are real-valued and $\phi(x_t) = x_t^2$, then $F$ represents the set of all asymptotically power limited signals with infinite duration.

**(2.6) Measurability condition (M), discrete block-memoryless input measures.** We consider the family $\mathcal{E} = \{E_s \subset X_0^s, s \in T_+\}$ of input constraints introduced in Paragraph 2.5 and define for any $s \in T_+$ the set

$$\mathfrak{A}(E_s) = \Big\{ \Big( \underset{k \in \mathbb{Z}}{\times} A_k, \bigotimes_{k \in \mathbb{Z}} \mathcal{A}_k \Big) : A_0 = \{\mathtt{a}_{[1]}, \ldots, \mathtt{a}_{[\mathtt{m}]}\} \subset E_s,$$
$$A_k = \langle A_0 \rangle_{ks}, \mathcal{A}_0 = 2^{A_0}, \mathcal{A}_k = \langle \mathcal{A}_0 \rangle_{ks} \Big\}, \quad (13)$$
$$\mathfrak{A}(\mathcal{E}) = \bigcup_{s \in T_+} \mathfrak{A}(E_s).$$

Thus $(A, \mathcal{A}) \in \mathfrak{A}(E_s)$ is a product measurable space with $A \subset E_s^* \subset X$ that is build from shifted copies of $(A_0, \mathcal{A}_0)$, where $A_0 \subset E_s \subset X_0^s$ contains finitely many signals in the time period $(0, s]$ satisfying the input constraint $E_s$ and $\mathcal{A}_0$ is the corresponding power set of $A_0$.

We say the channel $\kappa$ satisfies the measurability condition

(M) w.r.t. $\mathcal{E}$, if for all $(A, \mathcal{A}) \in \mathfrak{A}(\mathcal{E})$ and $B \in \mathcal{Y}$ it holds that $\kappa(\cdot, B)$ as function on $A$ is $\mathcal{A}$-measurable.

With this property $\kappa$ is a Markov kernel from $(A, \mathcal{A})$ to $(Y, \mathcal{Y})$ for all $(A, \mathcal{A}) \in \mathfrak{A}(\mathcal{E})$. According to the comment at the end of Remark 2.2, it is sufficient to verify that $\kappa(\cdot, G)$ as function on $A$ is $\mathcal{A}$-measurable for all $G \in \mathcal{G}$ and $(A, \mathcal{A}) \in \mathfrak{A}(\mathcal{E})$, where we can choose for example $\mathcal{G}$ to be equal to the families $\mathfrak{C}_Y$ or $\mathfrak{C}'_Y$ of cylinder sets given by

$$\mathfrak{C}_Y = \Big\{ [F] \subset Y : F \in \mathcal{Y}_u^v \text{ for some } u \leq v \in T \Big\},$$
$$\mathfrak{C}'_Y = \Big\{ \underset{t \in T}{\times} G_t \subset Y : G_t \in \mathcal{Y}_t, \quad (14)$$
$$\text{where only finitely many } G_t \neq Y_t \Big\}.$$

Indeed, the families $\mathfrak{C}_Y$ and $\mathfrak{C}'_Y$ generate $\mathcal{Y}$, i.e., $\mathcal{Y} = \sigma(\mathfrak{C}_Y)$ and $\mathcal{Y} = \sigma(\mathfrak{C}'_Y)$. Both are algebras and therefore in particular semirings, which are closed w.r.t. finite intersections.

The condition (M) fits to the measurability requirements in the proof of the coding theorem and converse theorem below. In particular, it allows to generate in a well-defined way the probability measure $\mu\kappa$ on $(A \times Y, \mathcal{A} \otimes \mathcal{Y})$ according to (1) using the channel $\kappa$ and a probability measure $\mu$ on $(A, \mathcal{A}) \in \mathfrak{A}(\mathcal{E})$. As will be shown in Lemma 2.8 the condition (M) is automatically satisfied by the channel memory properties needed anyway to prove the coding theorem. The measurability aspect of the channel is discussed in detail in Section III.B, in particular with regard to the commonly considered assumptions in the literature.

Subsequently, we are dealing with channel input probability measures of the following form. We define the sets of product probability measures

$$\mathcal{P}(E_s) = \Big\{ \bigotimes_{k \in \mathbb{Z}} \mu_k : \mu_0 = \sum_{i=1}^m p_i \delta_{\mathtt{a}_{[i]}}, \mathtt{a}_{[i]} \in E_s,$$
$$p_i > 0, \sum_{i=1}^m p_i = 1, \mu_k = \langle \mu_0 \rangle_{ks} \Big\}, \quad (15)$$
$$\mathcal{P}(\mathcal{E}) = \bigcup_{s \in T_+} \mathcal{P}(E_s), \quad (16)$$

where $\mu \in \mathcal{P}(E_s)$ is considered as measure on the product space $(A, \mathcal{A}) \in \mathfrak{A}(E_s)$ built from $A_0 = \{\mathtt{a}_{[1]}, \ldots, \mathtt{a}_{[\mathtt{m}]}\} \subset E_s$ as in (13). By $\delta_{\mathtt{a}_{[i]}}$ we denote the Dirac measure on $(A_0, \mathcal{A}_0)$ for the signal $\mathtt{a}_{[i]} \in A_0$. Further, we define the corresponding sets of probability spaces

$$\mathfrak{P}(E_s) = \Big\{ (A, \mathcal{A}, \mu) : (A, \mathcal{A}) \in \mathfrak{A}(E_s), \mu \in \mathcal{P}(E_s) \Big\}, \quad (17)$$
$$\mathfrak{P}(\mathcal{E}) = \bigcup_{s \in T_+} \mathfrak{P}(E_s). \quad (18)$$

We say a probability measure $\mu \in \mathcal{P}(E_s)$ is $s$-memoryless since w.r.t. the partitioned time axis into segments $(ks, (k+1)s], k \in \mathbb{Z}$, it represents a memoryless source. The alphabet of the source $\mu$ is finite and its elements satisfy the input constraint $E_s$. We call $\mathcal{P}(\mathcal{E})$ also the set of finite alphabet block-memoryless sources satisfying the constraints $\mathcal{E}$.

In the rest of the paper, $\mathcal{E} = \{E_s \subset X_0^s, s \in T_+\}$ denotes a family of input constraints and the notation introduced in Paragraphs 2.5 and 2.6 is used freely. The next definition specifies the channel properties required to derive the central coding theorem.

**(2.7) Definition (Channel properties: stationarity, ergodicity, causality, asymptotic input-memorylessness).**

(i) *Stationarity.* Given $s \in T$, the channel $\kappa$ is called $s$-stationary, if for any $x \in X$ and $B \in \mathcal{Y}$

$$\kappa(x, B) = \kappa(\langle x \rangle_s, \langle B \rangle_s)$$

holds. It is called stationary, if it is $s$-stationary for all $s \in T$.

(ii) *Ergodicity.* Assume that the channel $\kappa$ satisfies the measurability condition (M) w.r.t. the family $\mathcal{E} = \{E_s \subset X_0^s, s \in T_+\}$ of input constraints. For $s \in T_+$, an $s$-stationary channel $\kappa$ is called $s$-ergodic w.r.t. $\mathfrak{P}(E_s)$, if for any channel input probability space $(A, \mathcal{A}, \mu) \in \mathfrak{P}(E_s)$ the induced channel input-output probability measure $\mu\kappa$ on $(A \times Y, \mathcal{A} \otimes \mathcal{Y})$ is $s$-stationary and $s$-ergodic. A stationary channel $\kappa$ is called totally ergodic w.r.t. $\mathfrak{P}(\mathcal{E})$, if for all $s \in T_+$ it is $s$-ergodic

w. r. t. $\mathfrak{P}(E_s)$.

(iii) *Causality.* The channel $\kappa$ is called causal if for any $t \in T$, $B \in \mathcal{Y}_-^t$, and $x, \tilde{x} \in X$ coinciding on $(-\infty, t]$ we have

$$\kappa(x, [B]) = \kappa(\tilde{x}, [B]).$$

(iv) *Finite input memory.* The channel $\kappa$ has finite input memory if for all $s \in T$ there exists a $t_I(s) \in T_0$ such that for all $B \in \mathcal{Y}_s^+$ and $x, \tilde{x} \in X$ coinciding on $(s - t_I(s), \infty)$ we have

$$\kappa(x, [B]) = \kappa(\tilde{x}, [B]).$$

For given $s \in T$ we call the smallest possible $t_I(s)$ the input memory length at time $s$. If $t_I(s) = 0$ for all $s \in T$, then we say the channel is input-memoryless.

(v) *Asymptotic input-memorylessness.* The channel $\kappa$ is called asymptotically input-memoryless for the input signal set $X' \subset X$ if for any $\epsilon > 0$ and $s \in T$ there exists a $t_I(\epsilon, s) \in T_0$ such that for any $B \in \mathcal{Y}_s^+$ and $x, \tilde{x} \in X'$ coinciding on $(s - t_I(\epsilon, s), \infty)$ we have

$$\bigl|\kappa(x, [B]) - \kappa(\tilde{x}, [B])\bigr| < \epsilon.$$

Clearly, finite input memory implies asymptotic input-memorylessness. The next lemma is very useful because it mainly states that asymptotic input-memorylessness implies the measurability condition (M). Since asymptotic input-memorylessness is a form of continuous dependence on channel inputs remote in the past the lemma appears natural in some sense. However, it only holds due to the specific structure of the condition (M). A corresponding result for the measurability condition used in [3] does not hold. The lemma simplifies the verification of channel properties required in the general coding theorem. The proof is provided in Paragraph A.1 of Appendix A.

**(2.8) Lemma** *(Causal asymptotically input-memoryless channels satisfy (M)).* If the channel $\kappa$ is causal and asymptotically input-memoryless for the signal set $E' \subset X$ defined in (7) based on the family $\mathcal{E} = \{E_s \subset X_0^s, s \in T_+\}$ of input constraints, then $\kappa$ satisfies the measurability condition (M) w. r. t. $\mathcal{E}$.

The proof of the next preliminary result is given, e. g., in [24, p. 25] or [56, p. 124] for the case $T = \mathbb{Z}$. There is actually no difference in the derivations for $T = \mathbb{R}$.

**(2.9) Lemma** *(Stationarity of induced input-output probability measure).* Let $(A, \mathcal{A})$ with $A \subset X$ be a measurable space and assume that $\kappa(\cdot, B)$ as function on $A$ is $\mathcal{A}$-measurable for all sets $B \in \mathcal{Y}$, i.e., $\kappa$ is a Markov kernel from $(A, \mathcal{A})$ to $(Y, \mathcal{Y})$.

If $\kappa$ is an $s$-stationary (stationary) channel and $\mu$ is an $s$-stationary (stationary) channel input probability measure on $(A, \mathcal{A})$, then the induced input-output probability measure $\mu\kappa$ as defined by (1) and the corresponding output probability measure $\nu$ are $s$-stationary (stationary).

II.C. BLOCK CODES AND INFORMATION TRANSMISSION, CODING CAPACITY

For channels with time structure as defined in Definition 2.4 we introduce quantities relevant in connection with channel coding, such as block code, code rate, decoding error, maximal code size, coding capacity etc. The effect of channel input memory (also infinite) representing the potential impact of previously transmitted signals (codewords) on the current transmission is taken into account as well as possible restrictions on the signals allowed to send over the channel. Then we explain how a code is used in a communication system to transmit information, in particular to point out, that we consider codes that allow the repeated transmission of messages with the same reliability.

**(2.10) Definition (Block code with input constraint, code parameters).** Suppose the quantities $b \in T_+$, $E_b \subset X_0^b$, and $W = \{1, 2, \ldots, m\}$ for some $m \in \mathbb{N}$ are given. Let $u_i \in E_b$, $i \in W$, be pairwise distinct and let $B_i \in \mathcal{Y}_0^b$, $i \in W$, form a partition[4] of $Y_0^b$. Then

$$\mathcal{C}(b, E_b) = \bigl\{(u_i, B_i), i \in W\bigr\}$$

is called a (channel) block code for the message set $W$ satisfying the input constraint $E_b$, where $u_i$ is called codeword and $B_i$ decoding set for message $i \in W$. The code rate of $\mathcal{C}(b, E_b)$ is defined by

$$R_\mathcal{C} = \frac{1}{b} \log m,$$

where $m = |\mathcal{C}(b, E_b)|$ denotes the code size and $b$ the block length of $\mathcal{C}(b, E_b)$.

**(2.11) Definition (Decoding error probability, $(b, E_b, V, M, \epsilon)$-code).** Let $\mathcal{C}(b, E_b)$ be a block code as in Definition 2.10 and let

$$U_b^* = \bigtimes_{k \in \mathbb{Z}} \bigl\langle \{u_i, i \in W\} \bigr\rangle_{kb} \qquad (19)$$

be the set of all two-sided sequences of codewords. Suppose $V$ is a $b$-invariant set satisfying $U_b^* \subset V \subset X$ and assume that the channel $\kappa$ is $b$-stationary. The decoding error probability $\varrho(u_i, V)$ for codeword $u_i$ w. r. t. the set $V$ is defined by

$$\varrho(u_i, V) = \sup_{x \in [u_i] \cap V} \kappa(x, [B_i^c]),$$

where $B_i$ is the decoding set corresponding to the codeword $u_i$. The (maximal) decoding error probability $\varrho_{\max}(V)$ of the code $\mathcal{C}(b, E_b)$ is defined by

$$\varrho_{\max}(V) = \max_{i \in W} \varrho(u_i, V), \qquad (20)$$

where $W$ denotes the finite set of messages. The code $\mathcal{C}(b, E_b)$ is called a $(b, E_b, V, M, \epsilon)$-code if

$$|\mathcal{C}(b, E_b)| \geq M \qquad \text{and} \qquad \varrho_{\max}(V) \leq \epsilon.$$

---

[4]Assuming disjointness is essential here but the decoding sets do not have to fill up the whole space $Y_0^b$. However, the assumption is convenient and for our purposes it can always be established by merging the remaining part of $Y_0^b$ with one of the sets $B_i$.

**(2.12) Coding and decoding.** A block code $\mathcal{C}(b, E_b)$ is used in the following way to communicate messages from a set $W$ over a channel $\kappa$ with time structure. Let the operation start at time 0. To transmit message $i \in W$ codeword $u_i$ is sent within the time period $(0, b]$. If the received signal $v$ within the same time period lies in the decoding set $B_j$, then $v$ is decoded as message $j$. To transmit a sequence $(i_1, i_2, \ldots, i_n) \in W^n$ of $n$ messages, the described coding-decoding rule is applied $n$ times in succession. To transmit the $k$th message in the sequence, the time-shifted version $\langle u_{i_k} \rangle_{(k-1)b}$ of the codeword $u_{i_k}$ is sent within the time period $((k-1)b, kb]$. If the received signal $v$ within $((k-1)b, kb]$ lies in the time-shifted version $\langle B_j \rangle_{(k-1)b}$ of the decoding set $B_j$, then $v$ is decoded as message $j$. This message is then an estimate of $i_k$. The process of transmitting $n$ messages ends at time $nb$.

Figure 1 illustrates the coding-decoding process for a continuous-time channel with real input and output signals. The messages $2, 1, m, \ldots$ are encoded into the real functions $u_2, u_1, u_m, \ldots$ of duration $b$ taken from the set of codewords. The decoder identifies in each time interval of length $b$ the decoding set containing the received noisy signal, which is represented as a dot in the output space. In the example, the second message in the sequence is incorrectly decoded.

The code size $|\mathcal{C}(b, E_b)|$ is the number of different messages that can be communicated at most with the code. The block length $b$ specifies the time duration of transmitting a single message. The set $E_b$ represents the constraints imposed on the codewords as described in Paragraph 2.5. The code rate $R_\mathcal{C}$ characterizes the speed of transmission in [nat/channel use] if $T = \mathbb{Z}$ or in [nat/second] if $T = \mathbb{R}$.

For any $t \in T$ let $V_-^t$ and $V_t^+$ denote the images of the set $V$ w.r.t. the projections from $X$ onto $X_-^t$ and to $X_t^+$, respectively. Then $\varrho(u_i, V)$ is the probability that codeword $u_i$ — a signal in the time period $(0, b]$ — is not decoded as message $i$, no matter what signal from the set $V_-^0$ has been sent before and from the set $V_b^+$ has been sent afterwards. The assumption of a $b$-stationary channel $\kappa$ and a $b$-invariant set $V$ implies that $\varrho(u_i, V)$ is also equal to the probability that the shifted codeword $\langle u_i \rangle_{kb}$, $k \in \mathbb{N}$, is incorrectly decoded, regardless of past signals from the set $V_-^{kb}$ or future signals from the set $V_{(k+1)b}^+$. That means $\varrho(u_i, V)$ is the decoding error probability w.r.t. $V$ for codeword $u_i$ and for any version of $u_i$ shifted by a multiple of the block length $b$. The maximum of this decoding error probability over all codewords of $\mathcal{C}(b, E_b)$ is represented by $\varrho_{\max}(V)$. Subsequently, we restrict ourselves to situations, where we have the described shift-invariance of the decoding error. The introduced decoding error probability is based on [3, eq. (3)] and [12, eq. (2)]. It is defined in a worst case sense over the possible past input history and possible future inputs.

We require the set $V$ to contain the set $U_b^*$ of all sequences of codewords to make the decoding error at least robust w.r.t. past and future transmissions of codewords. Since it might be of interest to have robustness not only w.r.t. codewords, the definition allows to consider also larger sets of signals. As a typical example consider the set $E_b^* = \bigtimes_{k \in \mathbb{Z}} \langle E_b \rangle_{kb} \supset U_b^*$ already introduced in (5). For a transmission over time it is physically meaningful to assume the channel to be causal. Then the decoding error probability does only depend on past but not future inputs.

**(2.13) Maximal code size and coding capacity.** The maximal code size for a block length $b \in T_+$, an input constraint $E_b \subset X_0^b$, and a maximal decoding error probability of at most $\epsilon$ is given by

$$M^*(b, E_b, V, \epsilon) = \sup_{\mathcal{C}(b, E_b)} \{|\mathcal{C}(b, E_b)| : \varrho_{\max}(V) \leq \epsilon\},$$

where the set $V$ is specified as in Definition 2.11. Consequently, the corresponding maximal code rate is defined by

$$\frac{1}{b} \log M^*(b, E_b, V, \epsilon).$$

Given a family $\mathcal{E} = \{E_s \subset X_0^s, s \in T_+\}$ of input constraints, we are interested in the maximal code rate as a function of the block length $b$ for $b \to \infty$, i.e., the coding capacity

$$\overline{C} = \overline{C}(\mathcal{E}) = \inf_{\epsilon > 0} \limsup_{b \to \infty} \frac{1}{b} \log M^*(b, E_b, E_b^*, \epsilon).$$

In the definition of $\overline{C}$ we consider the maximal code rate w.r.t. the signal set $V = E_b^*$ introduced in (5), because this fits best to later derivations. $\overline{C}$ represents the supremum over all code rates $R$ such that for all $\epsilon > 0$ and $b_0 \in T_+$ there exists a $(b, E_b, E_b^*, e^{Rb}, \epsilon)$-code for some block length $b \geq b_0$. It follows that once we have found a "good" code, then there exist good codes for infinitely many larger block length. As the central result we show in Section III that if the channel $\kappa$ satisfies certain conditions – w.r.t. causality, stationarity, input memory, and output memory – then the operational quantity $\overline{C}$ is equal to the information rate capacity $C$ introduced in Section II.E. According to [58, p. 176] we call $\overline{C}$ optimistic coding capacity.

### II.D. INFORMATION MEASURES FOR GENERAL ALPHABETS

In order to fix notation but also to clearly point out the general scope of the results of this paper, we restate in this section definitions of information measures for random variables with abstract alphabet, which are necessary to formulate and derive the main results. Further, we provide two required lemmas on information rates and prove one of it since to the best of the authors knowledge a correct proof has seemingly not been published yet in the literature.

Throughout this section, the setting will be a given abstract probability space $(\Omega, \mathcal{F}, P)$. Unless stated otherwise all random variables are defined on this space and have values in arbitrary measurable spaces, specified only if required.

**(2.14) Definition ((Conditional) Mutual information, (conditional) entropy, information rate).** Assume that $\xi$ and $\eta$ are random variables with values in the measurable spaces $(X, \mathcal{X})$ and $(Y, \mathcal{Y})$. Then the mutual information $I(\xi; \eta)$ between $\xi$ and $\eta$ is defined by

$$I(\xi; \eta) = \sup \sum_{i=1}^{m} \sum_{j=1}^{n} P_{\xi, \eta}(A_i \times B_j) \log \frac{P_{\xi, \eta}(A_i \times B_j)}{P_\xi(A_i) P_\eta(B_j)},$$

where $P_\xi$, $P_\eta$, and $P_{\xi, \eta}$ denote the distribution of $\xi$, $\eta$, and $(\xi, \eta)$, respectively. Further assume that $\zeta$ is a random variable with values in the measurable spaces $(Z, \mathcal{Z})$. Then the conditional mutual information $I(\xi; \eta | \zeta)$ between $\xi$ and $\eta$ given $\zeta$ is defined by

$$I(\xi; \eta | \zeta) = \sup \sum_{i=1}^{m} \sum_{j=1}^{n} \mathrm{E}\bigg[ \mathrm{P}_{\xi, \eta | \zeta}(A_i \times B_j | \zeta) \cdot \log \frac{\mathrm{P}_{\xi, \eta | \zeta}(A_i \times B_j | \zeta)}{\mathrm{P}_{\xi | \zeta}(A_i | \zeta) \mathrm{P}_{\eta | \zeta}(B_j | \zeta)} \bigg],$$

where $\mathrm{P}_{\xi|\zeta}(A_i | z)$, $\mathrm{P}_{\eta|\zeta}(B_j | z)$, and $\mathrm{P}_{\xi,\eta|\zeta}(A_i \times B_j | z)$ denote for all $z \in Z$ the conditional probability of $\{\xi \in A_i\}$, $\{\eta \in B_j\}$, and $\{(\xi, \eta) \in A_i \times B_j\}$ given $\zeta = z$. The suprema are taken w.r.t. all finite partitions $\{A_1, A_2, \ldots, A_m\}$ of $X$ and $\{B_1, B_2, \ldots, B_n\}$ of $Y$ with $A_i \in \mathcal{X}$ and $B_j \in \mathcal{Y}$.

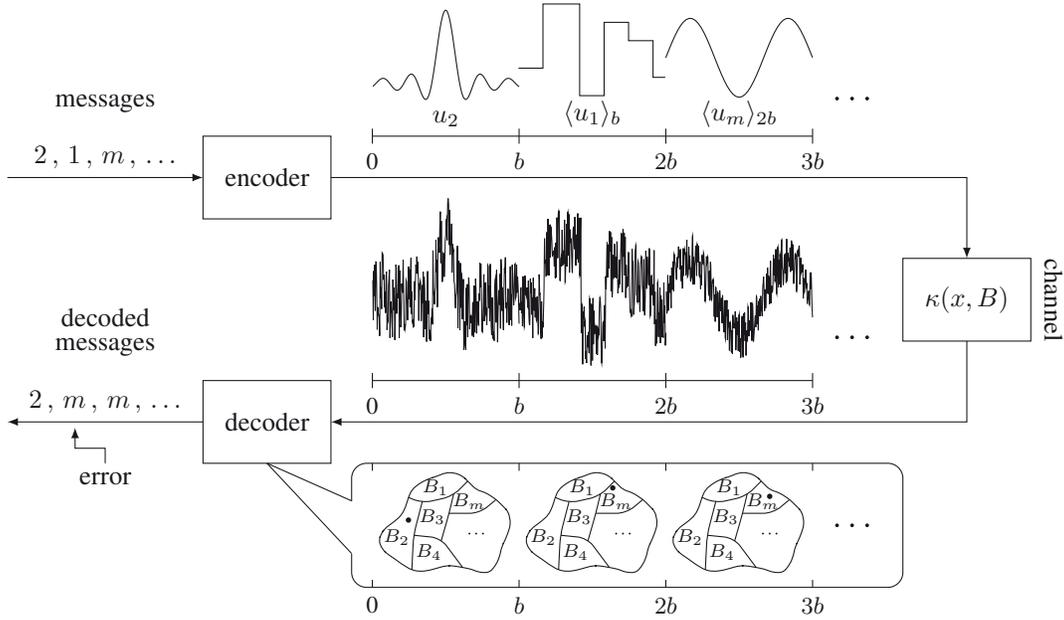

Fig. 1: Illustration of coded information transmission.

The entropy $H(\xi)$ of $\xi$ and the conditional entropy $H(\xi | \zeta)$ of $\xi$ given $\zeta$ are defined by

$$H(\xi) = I(\xi; \xi) \quad \text{and} \quad H(\xi | \zeta) = I(\xi; \xi | \zeta).$$

The (mutual) information rate $\bar{I}(\xi; \eta)$ between the random processes $\xi = \{\xi_t, t \in T\}$ and $\eta = \{\eta_t, t \in T\}$ is defined by

$$\bar{I}(\xi; \eta) = \lim_{s \to \infty} \frac{1}{s} I(\xi_0^s; \eta_0^s),$$

if the limit exists.

**(2.15) Remark.** The definition of mutual information for random variables with abstract alphabet is taken from [59, Sec. 2.1]. The definition of conditional mutual information is a version of that given by Wyner [60]. It is a completely general and direct definition and has the advantage that it generalizes Dobrushin's indirect and restricted definition of conditional mutual information considered in [59] to cases, for which certain regular conditional probabilities do not exist [59, see translator's remarks to Ch. 3]. A standard reference for information measures in general form is [59], where the properties used in subsequent derivations can be found (see [59, Ch. 2 and 3]). However, several of these properties require proofs different from those in [59] if Wyner's generalized form of conditional mutual information is employed (like in the abstract context of this paper). See [60] and [5, Rmk. 4.8] in this regard. As in [59, eq. (5.4.1)] the (mutual) information rate is defined in the most basic and natural way, which is sufficient for the purposes of this paper.

**(2.16) Lemma (*Information rate for stationary random sequences, one being independent*).** Let $\xi = \{\xi_k, k \in \mathbb{Z}\}$ be a sequence of independent and $\eta = \{\eta_k, k \in \mathbb{Z}\}$ be a sequence of arbitrary random variables such that the pair sequence $\{(\xi_k, \eta_k), k \in \mathbb{Z}\}$ is stationary. Then

(i) the sequence $\{n^{-1} I(\xi_0^n; \eta_0^n), n \in \mathbb{N}\}$ is monotonically increasing,
(ii) the sequence $\{I(\xi_0^n; \eta_0^n), n \in \mathbb{N}\}$ is superadditive, i.e., for all $l, m \in \mathbb{N}$ we have

$$I(\xi_0^l; \eta_0^l) + I(\xi_0^m; \eta_0^m) \leq I(\xi_0^{l+m}; \eta_0^{l+m}),$$

and (i) as well as (ii) imply that the information rate $\bar{I}(\xi; \eta)$ exists and is given by

$$\bar{I}(\xi; \eta) = \sup_{n \geq 1} \frac{1}{n} I(\xi_0^n; \eta_0^n).$$

**(2.17) Lemma (*Information rate for stationary sequences, one with finite alphabet*).** Let $\xi = \{\xi_k, k \in \mathbb{Z}\}$ and $\eta = \{\eta_k, k \in \mathbb{Z}\}$ be random sequences such that the pair sequence $\{(\xi_k, \eta_k), k \in \mathbb{Z}\}$ is stationary. Further assume that the random variables $\xi_k$ have a finite alphabet. Then the information rate $\bar{I}(\xi; \eta)$ exists and is given by

$$\bar{I}(\xi; \eta) = I(\xi_1; \eta | \xi_-^0). \tag{21}$$

In particular, $\bar{I}(\xi; \eta)$ is finite.

**(2.18) Remark.** Lemma 2.16 is a useful basic result, which is employed in the proof of the central coding theorem (converse) for abstract channels with time structure. It is a special case of a more general version shown in [5, Lem. 4.13]. Note, that the monotonicity result is used also in [3] to prove a coding theorem for continuous-time channels. Further, it is employed in [61] and [62]. However, the proof given in [3, Appendix II] regarding the monotonicity is not correct as shown with a counterexample in Appendix C. Since a complete proof has seemingly not been published yet in the literature it is given in Paragraph A.2 of Appendix A.

In Lemma 2.17 the proof of the existence of $\bar{I}(\xi;\eta)$ and of the equality (21) is identical to the proof of the first part of [59, Thm. 6.1.1]. In [59] it is additionally assumed that also the random variables $\eta_k$, $k \in \mathbb{Z}$, have a finite alphabet, which is actually not relevant. Since we have

$$I(\xi_1;\eta|\xi_-^0) \leq H(\xi_1|\xi_-^0) \leq H(\xi_1),$$

where $H(\xi_1)$ is finite, we obtain that $\bar{I}(\xi;\eta)$ is finite.

## II.E. Information Rate Capacity

We define the information rate capacity of a channel $\kappa$ with time structure as introduced in Definition 2.4 for the family $\mathcal{E} = \{E_s \subset X_0^s, s \in T_+\}$ of input constraints.

**(2.19) Definition (Information rate capacity).** Given the channel $\kappa$ satisfies the measurability condition (M) w.r.t. to $\mathcal{E}$, then we define the information rate capacity of $\kappa$ for the family $\mathcal{E}$ of constraints by

$$\begin{aligned} C &= C(\mathcal{E}) = \limsup_{s \to \infty} \frac{1}{s} C_s, \\ C_s &= C_s(E_s) = \sup_{\mu \in \mathcal{P}(E_s)} I\big(\xi_0^s;\eta_0^s\big), \end{aligned} \quad (22)$$

where $\mu \in \mathcal{P}(E_s)$ is considered as probability measure on the corresponding measurable space $(A, \mathcal{A}) \in \mathfrak{A}(E_s)$ as in (15). The mutual information $I\big(\xi_0^s;\eta_0^s\big)$ is computed between the projections $\xi_0^s$ and $\eta_0^s$ from $A \times Y$ onto $A_0 \subset X_0^s$ and $Y_0^s$, regarded as random variables on the channel input-output probability space $(A \times Y, \mathcal{A} \otimes \mathcal{Y}, \mu\kappa)$.

**(2.20) Remark.** To define the information rate capacity of a channel with time structure the time index set $T$ is at first partitioned into segments of size $s \in T_+$. The quantity $C_s$ is the supremum of the mutual information between the channel input and output within the time period $(0, s]$, where the supremum is taken w.r.t. all $\mu \in \mathcal{P}(E_s)$, i.e., w.r.t. all finite alphabet $s$-memoryless input sources satisfying the constraint $E_s$. Normalizing $C_s$ by the time duration $s$ and taking the limit superior w.r.t. $s$ defines the information rate capacity $C$ in [nat/channel use] if $T = \mathbb{Z}$ or in [nat/second] if $T = \mathbb{R}$. It will be shown that – under suitable conditions – $C$ is equal to the (optimistic) coding capacity, even though $C$ is obtained by optimizing "only" w.r.t. a restricted class of finite alphabet block-memoryless input sources.

In the next lemma, different representations of the information rate capacity are given. The first is useful, e.g., to prove the coding theorem (converse) for stationary channels. The second relates Definition 2.19 to a mathematically useful process definition. For details please refer to the discussion in [5, Rmk. 5.4]. A proof of the lemma is given in Paragraph A.3 of Appendix A and is mainly based on Lemma 2.16.

**(2.21) Lemma (*Information rate capacity of stationary channels*).** *Consider the information rate capacity $C = \limsup_{s\to\infty} C_s/s$ introduced in Definition 2.19 and additionally assume that the channel $\kappa$ is stationary and the family $\mathcal{E}$ of input constraints satisfies the regularity condition (R) in (8). Then we have the identities*

$$C = \sup_{s \in T_+} \frac{1}{s} C_s, \quad (23)$$

$$C = \sup_{\mu \in \mathcal{P}(\mathcal{E})} \bar{I}(\mu), \qquad \bar{I}(\mu) = \lim_{n \to \infty} \frac{1}{ns} I\big(\xi_0^{ns};\eta_0^{ns}\big), \quad (24)$$

*where $\mathcal{P}(\mathcal{E})$ is defined in (16) and $s \in T_+$ in the definition of $\bar{I}(\mu)$ is the size of the segments of the time axis $T$ belonging to $\mu \in \mathcal{P}(E_s)$. Further, $\xi_0^{ns} = \big(\xi_0^s, \xi_s^{2s}, \ldots, \xi_{(n-1)s}^{ns}\big)$ and $\eta_0^{ns} = \big(\eta_0^s, \eta_s^{2s}, \ldots, \eta_{(n-1)s}^{ns}\big)$, where $\xi_{ks}^{(k+1)s}$ and $\eta_{ks}^{(k+1)s}$ denote the projections from $A \times Y$ onto $A_k = \langle A_0 \rangle_{ks_0} \subset X_{ks}^{(k+1)s}$ and $Y_{ks}^{(k+1)s}$, respectively.*

## II.F. Proving Achievability and Converse

**(2.22) Lemma (*Feinstein's lemma*).** *Suppose $\mathrm{P}_1$ is a probability measure on the measurable space $(\Omega_1, \mathcal{F}_1)$ and $K$ is a Markov kernel from $(\Omega_1, \mathcal{F}_1)$ to the measurable space $(\Omega_2, \mathcal{F}_2)$. Further, let $\mathrm{P}$ denote the probability measure induced by $\mathrm{P}_1$ and $K$, i.e., the measure on the product space $(\Omega_1 \times \Omega_2, \mathcal{F}_1 \otimes \mathcal{F}_2)$ defined by*

$$\mathrm{P}(F_1 \times F_2) = \int_{F_1} K(\omega_1, F_2) \, \mathrm{dP}_1(\omega_1), \quad F_1 \in \mathcal{F}_1, F_2 \in \mathcal{F}_2.$$

*We assume that $\mathrm{P}$ has a $\mathrm{P}_1 \otimes \mathrm{P}_2$-density, say $f$, where $\mathrm{P}_2$ denotes the marginal measure of $\mathrm{P}$ on $(\Omega_2, \mathcal{F}_2)$. Let $\gamma \in \mathbb{R}$ and $m \in \mathbb{N}$ be arbitrary and consider the set*

$$G_\gamma = \{f > e^\gamma\} = \{(\omega_1, \omega_2) \in \Omega_1 \times \Omega_2 : f(\omega_1, \omega_2) > e^\gamma\}.$$

*Then there exist elements $a_1, a_2, \ldots, a_m \in \Omega_1$ and a partition $\{B_1, B_2, \ldots, B_m\}$ of $\Omega_2$ with $B_i \in \mathcal{F}_2$ such that*

$$K(a_i, B_i^c) < m e^{-\gamma} + \mathrm{P}(G_\gamma^c) \quad (25)$$

*holds for all $i = 1, 2, \ldots, m$.*

**(2.23) Remark.** Feinstein's lemma is a tool to prove coding theorems for channels with memory in a transparent way. This version of the lemma is what we require in the proof of the coding theorem. The original lemma is published in [63]. A generalization is provided in [5, Lem. 8.1] together with historical notes and a discussion on various forms of the lemma in [5, Rmk. 8.2]. A proof of the version given above can be found in [24, Sec. 14.2].

To prove a coding theorem using Feinstein's lemma it is necessary to control the upper bound of the (conditional) probability in (25), where the probability will correspond to the decoding error. This is possible with the help of the ergodic theorem of information theory stated next for a quite simple special case, which turns out to be sufficient in the context of this paper.

**(2.24) Theorem** *(Ergodic theorem of information theory for stationary sequences, one with finite alphabet).* *Assume that $\xi = \{\xi_k, k \in \mathbb{Z}\}$ and $\eta = \{\eta_k, k \in \mathbb{Z}\}$ are random sequences on the probability space $(\Omega, \mathcal{F}, \mathrm{P})$. Further assume that the pair sequence $\{(\xi_k, \eta_k), k \in \mathbb{Z}\}$ is stationary and ergodic and that the random variables $\xi_k$ have a finite alphabet. Then for any $\epsilon > 0$ we have*

$$\lim_{n \to \infty} \mathrm{P}\left(\left|\frac{1}{n} \log f^{(n)}(\xi_0^n, \eta_0^n) - \bar{I}(\xi; \eta)\right| \geq \epsilon\right) = 0, \quad (26)$$

*where $f^{(n)}$ denotes the $\mathrm{P}_{\xi_0^n} \otimes \mathrm{P}_{\eta_0^n}$-density of $\mathrm{P}_{\xi_0^n, \eta_0^n}$. By $\mathrm{P}_{\xi_0^n, \eta_0^n}$, $\mathrm{P}_{\xi_0^n}$, and $\mathrm{P}_{\eta_0^n}$ we denote the distribution of $(\xi_0^n, \eta_0^n)$, $\xi_0^n$, and $\eta_0^n$, respectively.*

**(2.25) Remark.** Due to the finite alphabet of the random variable $\xi_k$ for all $k \in \mathbb{N}$ the mutual information $I(\xi_0^n; \eta_0^n)$ is finite for all $n \in \mathbb{N}$. Further, the information rate $\bar{I}(\xi; \eta)$ exists and is finite (see Lemma 2.17). Thus, the density $f^{(n)}$ exists for all $n \in \mathbb{N}$ due to the integral form of mutual information (see [59, Thm. 2.1.2]) such that (26) is well-defined.

This version of the ergodic theorem is precisely what we require to prove the coding theorem. The proof of Theorem 2.24 is identical to the proof of [59, Th. 6.4.1], where it is additionally assumed that also the random variables $\eta_k, k \in \mathbb{N}$, have a finite alphabet. That we can follow the proof given there prerequires Dobrushin's version of the involved conditional mutual information (see Remark 2.15) exists, including the corresponding so called conditional information densities. Indeed, since $f^{(n)}$ exists for all $n \in \mathbb{N}$, the comment in [59, p. 56, first paragraph] implies the existence of the conditional mutual information in the sense of Dobrushin. These information quantities are all finite due to the finite alphabet of the random variables $\xi_k, k \in \mathbb{N}$, such that the existences of the corresponding conditional information densities is guaranteed according to [59, p. 29].

A generalized version of the ergodic theorem is given in [59, Th. 8.2.1] or [5, Th. 8.4]. Historical notes and a discussion on various forms of the ergodic theorem is given in [5, Rmk. 8.5].

Fano's inequality, originally published in [64], is a standard tool to prove a weak converse of a coding theorem. See [65, p. 81] or [34, p. 78–79] for a proof of this result.

**(2.26) Lemma** *(Fano's inequality).* *Let $\xi$ and $\eta$ be discrete random variables on the probability space $(\Omega, \mathcal{F}, \mathrm{P})$ with common finite alphabet of size $m > 1$. Then the conditional entropy $H(\xi|\eta)$ of $\xi$ given $\eta$ satisfies the inequality*

$$h_m(p_e) \geq H(\xi|\eta),$$

*where $p_e = \mathrm{P}(\xi \neq \eta)$ and the function $h_m$ is defined for integers $m > 1$ and all $x \in [0, 1]$ by*

$$h_m(x) = x \log(m-1) - x \log x - (1-x) \log(1-x). \quad (27)$$

## III. CODING THEOREM AND WEAK CONVERSE

For an abstract channel $\kappa$ with time structure as introduced in Definition 2.4 with input product space $X$ and output product measurable space $(Y, \mathcal{Y})$ we prove under quite general conditions a block coding theorem and a converse taking general input constraints into account, represented by the family $\mathcal{E} = \{E_s \subset X_0^s, s \in T_+\}$ of constraint sets.

### III.A. THEOREMS AND PROOFS

**(3.1) Theorem** *(Coding theorem).* *Suppose the channel $\kappa$ is stationary, causal, asymptotically input-memoryless for the signal set $E'$, and totally ergodic w.r.t. $\mathfrak{P}(\mathcal{E})$, with $E'$ and $\mathfrak{P}(\mathcal{E})$ as defined in (7) and (18). Further assume that the family $\mathcal{E}$ of input constraints satisfies the regularity condition (R) in (8).*

*(i) If the information rate capacity $C$ for the constraints $\mathcal{E}$ is finite, then for any $\rho \in (0, C)$, $\epsilon \in (0, 1)$, and $b_0 \in T_+$ there exists a $(b, E_b, E', \mathrm{e}^{(C-\rho)b}, \epsilon)$-code for some block length $b \geq b_0$.*

*(ii) If $C$ is infinite, then for any $R > 0$, $\epsilon \in (0, 1)$, and $b_0 \in T_+$ there exists a $(b, E_b, E', \mathrm{e}^{Rb}, \epsilon)$-code for some block length $b \geq b_0$.*

*Proof:* (i) *Setup.* Let us fix some $\epsilon \in (0, 1)$. If $C < \infty$ we fix some $\rho \in (0, C)$ and put $R = C - \rho$. If $C = \infty$ then let $R > 0$ be arbitrary. Due to the definition of the information rate capacity given in Definition 2.19 there exists in both cases an $s_0 \in T_+$ and a channel input probability space $(A, \mathcal{A}, \mu) \in \mathfrak{P}(E_{s_0})$ such that

$$R < \frac{1}{s_0} I(\xi_0^{s_0}; \eta_0^{s_0}). \quad (28)$$

Let $\alpha = \{\alpha_k, k \in \mathbb{Z}\}$ and $\beta = \{\beta_k, k \in \mathbb{Z}\}$ denote the sequences of projections on the channel input-output probability space $(A \times Y, \mathcal{A} \otimes \mathcal{Y}, \mu\kappa)$ as introduced in (68) in the proof of Lemma 2.21. Since the channel $\kappa$ is assumed causal and asymptotically input-memoryless for the signal set $E'$ it satisfies the measurability condition (M) due to Lemma 2.8. Therefore, the probability measure $\mu\kappa$ is well-defined. Further, let $\mu_0$ denote the probability measure on $(A_0, \mathcal{A}_0) = (\{\mathsf{a}_{[1]}, \mathsf{a}_{[2]}, \ldots, \mathsf{a}_{[M]}\}, 2^{A_0})$ as given in (15) from which the product measure $\mu \in \mathfrak{P}(E_{s_0})$ is built.

The channel $\kappa$ is assumed to be asymptotically input-memoryless for the set $E'$. Due to (2.7.v) this implies there exists a minimal $l_0 \in \mathbb{N}$, such that for any $B \in \mathcal{Y}_0^+$ and $x, \tilde{x} \in E'$ coinciding on $(-l_0 s_0, \infty)$ we have

$$\left|\kappa(x, [B]) - \kappa(\tilde{x}, [B])\right| < \frac{\epsilon}{2}. \quad (29)$$

It follows together with the stationarity of $\kappa$, the equality $\mathcal{Y}_{l_0 s_0}^+ = \langle \mathcal{Y}_0^+ \rangle_{l_0 s_0}$, and the shift-invariance of the set $E'$ that (29) holds for any $B \in \mathcal{Y}_{l_0 s_0}^+$ and $x, \tilde{x} \in E'$ coinciding on $(0, \infty)$.

(ii) *Ergodic theorem.* The channel $\kappa$ is assumed to be stationary so that it is in particular $s_0$-stationary according to (2.7.i). Together with the $s_0$-stationarity of $\mu$ we obtain from Lemma 2.9 the $s_0$-stationarity of the channel input-output probability measure $\mu\kappa$. Further, the channel $\kappa$ is assumed to be totally ergodic w.r.t. $\mathfrak{P}(\mathcal{E})$, thus it is $s_0$-ergodic w.r.t. $\mathfrak{P}(E_{s_0})$. According to (2.7.ii) we obtain together with $\mu \in \mathfrak{P}(E_{s_0})$ that $\mu\kappa$ is $s_0$-ergodic. Therefore, the pair sequence $\{(\alpha_k, \beta_k), k \in \mathbb{Z}\}$ of projections is stationary and ergodic.

Furthermore, $\{\alpha_k, k \in \mathbb{Z}\}$ is a sequence of discrete random variables with finite alphabet of size $M$, since $\alpha_{k+1}$ has values in $A_k = \langle A_0 \rangle_{ks_0}$. Thus, the information rate $\bar{I}(\alpha; \beta)$ exists and is finite due to Lemma 2.17. Moreover, we can apply Theorem 2.24 and obtain for any $\epsilon^* > 0$

$$\lim_{n \to \infty} \mu\kappa\left(f^{(n)}(\alpha_0^n, \beta_0^n) \leq e^{n(\bar{I}(\alpha;\beta) - \epsilon^*)}\right) = 0, \quad (30)$$

where $f^{(n)}$ denotes the density of the distribution of $(\alpha_0^n, \beta_0^n)$ w.r.t. the product of the distributions of $\alpha_0^n$ and $\beta_0^n$. This density exists for all $n \in \mathbb{N}$ since $\bar{I}(\alpha; \beta)$ is finite (see beginning of Remark 2.25).

(iii) *Code size.* Let us fix

$$\epsilon^* = \frac{1}{3}(\bar{I}(\alpha; \beta) - Rs_0). \quad (31)$$

As in in the proof of Lemma 2.21 in Paragraph A.3 we obtain from $\mu \in \mathcal{P}(E_{s_0})$ and $\kappa$ being stationary that Lemma 2.16 can be applied to the sequences $\alpha$ and $\beta$, which yields the inequalities

$$\frac{1}{s_0} I(\alpha_1; \beta_1) \leq \frac{1}{ns_0} I(\alpha_0^n; \beta_0^n) \leq \frac{1}{s_0} \bar{I}(\alpha; \beta) \quad (32)$$

for any $n \in \mathbb{N}$. From (28) and (32) we therefore obtain that $\epsilon^*$ is positive. Further, due to (30) we can choose a sufficiently large $n_0 \in \mathbb{N}$, such that

$$e^{-n_0 \epsilon^*} < \frac{\epsilon}{4} \quad (33)$$

$$Rl_0 s_0 + 1 < n_0 \epsilon^* \quad (34)$$

$$\mu\kappa\left((\alpha_0^{n_0}, \beta_0^{n_0}) \in G_\gamma^c\right) < \frac{\epsilon}{4} \quad (35)$$

hold simultaneously, where

$$G_\gamma = \{f^{(n_0)} > e^\gamma\} \quad \text{and} \quad \gamma = n_0(\bar{I}(\alpha; \beta) - \epsilon^*). \quad (36)$$

We continue by choosing an $m \in \mathbb{N}$ that satisfies

$$e^{R(n_0 + l_0)s_0} < m < e^{Rn_0 s_0 + n_0 \epsilon^*}, \quad (37)$$

which is possible due to (34). From (31), (33), and the right-hand side of (37) we obtain

$$me^{-\gamma} = e^{-2n_0 \epsilon^* + (\log m - Rn_0 s_0)} < e^{-n_0 \epsilon^*} < \frac{\epsilon}{4}. \quad (38)$$

(iv) *Feinstein's lemma.* Let us put $t_0 = n_0 s_0$. Based on the probability space $(A_0, \mathcal{A}_0, \mu_0)$ that generates the channel input probability space $(A, \mathcal{A}, \mu) \in \mathfrak{P}(E_{s_0})$ we define the (sub-) product measures

$$\mu_- = \bigotimes_{k=-1}^{-\infty} \mu_k \quad \text{and} \quad \mu'_0 = \bigotimes_{k=0}^{n_0 - 1} \mu_k \quad (39)$$

on the (sub-) measurable spaces

$$(A_-, \mathcal{A}_-) = \left(\underset{k=-1}{\overset{-\infty}{\times}} A_k, \underset{k=-1}{\overset{-\infty}{\bigotimes}} \mathcal{A}_k\right) \quad \text{and}$$

$$(A'_0, \mathcal{A}'_0) = \left(\underset{k=0}{\overset{n_0-1}{\times}} A_k, \underset{k=0}{\overset{n_0-1}{\bigotimes}} \mathcal{A}_k\right),$$

respectively, where for all $k \in \mathbb{Z}$ we have $(A_k, \mathcal{A}_k, \mu_k) = (\langle A_0 \rangle_{ks_0}, \langle \mathcal{A}_0 \rangle_{ks_0}, \langle \mu_0 \rangle_{ks_0})$.

Since the channel $\kappa$ is assumed causal we have for all $B \in \mathcal{Y}_0^{t_0}$ that $\kappa(a, [B])$ does only depend on coordinates of $a = \{a_k, k \in \mathbb{Z}\} \in A$ with index $k < n_0$, i.e., $\kappa(\cdot, [B])$ is in fact not a function on $A$ but just on $A_- \times A'_0$. Therefore, we can write $\kappa((a_-, a'_0), [B])$ with $(a_-, a'_0) \in A_- \times A'_0$ instead of $\kappa(a, [B])$. Further, since $\kappa$ is assumed causal and asymptotically input-memoryless for the signal set $E'$ it satisfies the measurability condition (M) due to Lemma 2.8. Therefore, $\kappa$ as function on $(A_- \times A'_0) \times [\mathcal{Y}_0^{t_0}]$ is a Markov kernel from $(A_- \times A'_0, \mathcal{A}_- \otimes \mathcal{A}'_0)$ to $(Y, [\mathcal{Y}_0^{t_0}])$ and we can define for any $a'_0 \in A'_0$ and $B \in \mathcal{Y}_0^{t_0}$

$$\bar{\kappa}(a'_0, B) = \int_{A_-} \kappa((a_-, a'_0), [B]) \, d\mu_-(a_-). \quad (40)$$

From basic measure-theoretic results it follows that $\bar{\kappa}$ is a Markov kernel from $(A'_0, \mathcal{A}'_0)$ to $(Y_0^{t_0}, \mathcal{Y}_0^{t_0})$. Together with $\mu'_0$ it induces a probability measure on $(A'_0 \times Y_0^{t_0}, \mathcal{A}'_0 \otimes \mathcal{Y}_0^{t_0})$, say $\mu\kappa'_0$, given for any $F \in \mathcal{A}'_0$ and $G \in \mathcal{Y}_0^{t_0}$ by

$$\mu\kappa'_0(F \times G) = \int_F \bar{\kappa}(a'_0, G) \, d\mu'_0(a'_0).$$

From the definition of $\mu\kappa$, the product structure of $\mu \in \mathcal{P}(E_{s_0})$, the causality of $\kappa$, the definition of $\bar{\kappa}$, and basic measure-theoretic results (Fubini's theorem) we obtain that $\mu\kappa'_0$ is the distribution of $(\alpha_0^{n_0}, \beta_0^{n_0})$, since it is equal to the marginal measure of $\mu\kappa$ on $(A'_0 \times Y_0^{t_0}, \mathcal{A}'_0 \otimes \mathcal{Y}_0^{t_0})$. Furthermore, $\mu'_0$ is the distribution of $\alpha_0^{n_0}$ and $\nu'_0$ is the distribution of $\beta_0^{n_0}$, where $\nu'_0$ denotes the marginal measure of $\mu\kappa'_0$ on $(Y_0^{t_0}, \mathcal{Y}_0^{t_0})$. It follows that the function $f^{(n_0)}$ used to define the set $G_\gamma$ in (36) is the $\mu'_0 \otimes \nu'_0$-density of $\mu\kappa'_0$ and

$$\mu\kappa'_0(G_\gamma^c) = \mu\kappa\left((\alpha_0^{n_0}, \beta_0^{n_0}) \in G_\gamma^c\right). \quad (41)$$

Applying Feinstein's lemma given in Lemma 2.22 we obtain that there exist

$$u'_1, u'_2, \ldots, u'_m \in A'_0 \quad \text{and} \quad \hat{B}_1, \hat{B}_2, \ldots, \hat{B}_m \in \mathcal{Y}_0^{t_0}$$

such that

$$\bar{\kappa}(u'_i, \hat{B}_i^c) < me^{-\gamma} + \mu\kappa'_0(G_\gamma^c) < \frac{\epsilon}{2}, \quad (42)$$

holds for all $i \in \{1, 2, \ldots, m\}$, where $\gamma$ and $m$ are chosen as in (36) and (37). The upper bound of $\epsilon/2$ is obtained from (35), (38), and (41).

(v) *Code construction.* The channel $\kappa$ satisfies the measurability condition (M) w.r.t. $\mathcal{E}$ so that $\kappa((\cdot, u'_i), [\hat{B}_i^c])$ is a random variable on the probability space $(A_-, \mathcal{A}_-, \mu_-)$ with expectation $\bar{\kappa}(u'_i, \hat{B}_i^c)$. Due to the causality, $\kappa(\cdot, [\hat{B}_i^c])$ is considered here again as function on $A_- \times A'_0$. If the expectation of a random variable is less than some constant, then the probability that the random variable has values less than this constant is positive. Therefore, we obtain from (42)

$$\mu_-\left(\kappa\left((\cdot, u'_i), [\hat{B}_i^c]\right) < \epsilon/2\right) > 0.$$

This implies that for all $i \in \{1, 2, \ldots, m\}$ there exists a $\hat{u}_i \in A_- \times A'_0$ coinciding on $(0, n_0 s_0]$ with $u'_i$ such that

$$\kappa(\hat{u}_i, [\hat{B}_i^c]) < \frac{\epsilon}{2}.$$

All $\hat{u}_i$ can be chosen $\hat{u}_i \in A$ as well due to the causality of the channel $\kappa$. The stationarity of $\kappa$ then implies

$$\kappa\big(\hat{u}_i, [\hat{B}_i^c]\big) = \kappa\big(\langle\hat{u}_i\rangle_{l_0 s_0}, \langle[\hat{B}_i]\rangle_{l_0 s_0}\big) < \frac{\epsilon}{2}. \quad (43)$$

Let $u_i$ denote the projection of $\langle\hat{u}_i\rangle_{l_0 s_0}$ onto $X_0^b$ and $B_i$ the projection of $\langle[\hat{B}_i]\rangle_{l_0 s_0}$ onto $Y_0^b$, where $b$ is given by

$$b = (n_0 + l_0) s_0.$$

The code

$$\mathcal{C} = \{(u_i, B_i), i \in \{1, 2, \ldots, m\}\}$$

has a block length $b$ and a code size $m > e^{Rb}$ due to (37). Each decoding set satisfies $B_i \in \mathcal{Y}_0^b$ and $[B_i] = \langle[\hat{B}_i]\rangle_{l_0 s_0} \in [\mathcal{Y}_{l_0 s_0}^b]$. Because $u_i \in \times_{k=0}^{n_0+l_0-1} \langle A_0\rangle_{k s_0}$ and $A_0 \subset E_{s_0}$ we have for each codeword $u_i \in E_b$, since we assume that the family $\mathcal{E}$ of input constraints satisfies the regularity condition (R) in (8). Furthermore, we assume that the channel $\kappa$ is causal and asymptotically input-memoryless for the set $E'$ so that (43) and the comment below (29) imply for all $i \in \{1, 2, \ldots, m\}$

$$\kappa\big(x_i, [B_i^c]\big) < \epsilon,$$

for all $x_i \in E'$ coinciding on $(0, b)$ with $u_i$. The maximal decoding error $\varrho_{\max}(E')$ is therefore bounded by $\epsilon$. Thus, $\mathcal{C}$ is a $(b, E_b, E', e^{Rb}, \epsilon)$-code, which completes the proof. ∎

**(3.2) Theorem** *(Weak converse). Assume that the channel $\kappa$ is stationary, satisfies the measurability condition (M) w.r.t. the family $\mathcal{E}$ of input constraints, and $\mathcal{E}$ satisfies the regularity condition (R) in (8).*

*If the information rate capacity $C$ of $\kappa$ for the constraints $\mathcal{E}$ is finite, then for any $\rho > 0$ there exists a constant $\epsilon^* = \epsilon^*(\rho, C) \in (0, 1)$, such that for any $\epsilon < \epsilon^*$ and block length $b \in T_+$ there does not exist a $(b, E_b, U_b^*, e^{(C+\rho)b}, \epsilon)$-code, where $U_b^*$ is defined in (19).*

*Proof:* (i) *Setup.* Let us fix some $\rho > 0$, a block length $b \in T_+$, and an integer $m \geq e^{(C+\rho)b}$. Further, let

$$\mathcal{C}(b, E_b) = \{(u_i, B_i), i \in \{1, 2, \ldots, m\}\}$$

be some block code satisfying the input constraint $E_b$. We define the probability space $(A_0, \mathcal{A}_0, \mu_0)$ with

$$A_0 = \{u_1, u_2, \ldots, u_m\} \subset E_b \subset X_0^b,$$
$$\mathcal{A}_0 = 2^{A_0}, \quad \mu_0 = \sum_{i=1}^m p_i \delta_{u_i}, \quad (44)$$

where we choose without loss of generality $p_i = 1/m$. Further, we define the product measures

$$\mu_- = \bigotimes_{k=-1}^{-\infty} \mu_k \quad \text{and} \quad \mu_+ = \bigotimes_{k=1}^{\infty} \mu_k \quad (45)$$

on the measurable spaces

$$(A_-, \mathcal{A}_-) = \Big(\mathop{\times}_{k=-1}^{-\infty} A_k, \bigotimes_{k=-1}^{-\infty} \mathcal{A}_k\Big) \quad \text{and}$$
$$(A_+, \mathcal{A}_+) = \Big(\mathop{\times}_{k=1}^{\infty} A_k, \bigotimes_{k=1}^{\infty} \mathcal{A}_k\Big),$$

respectively, where $(A_k, \mathcal{A}_k, \mu_k) = \big(\langle A_0\rangle_{kb}, \langle\mathcal{A}_0\rangle_{kb}, \langle\mu_0\rangle_{kb}\big)$ for all $k \in \mathbb{Z}$. Subsequently, we consider

$$(A, \mathcal{A}, \mu) = \\ \big(A_- \times A_0 \times A_+, \mathcal{A}_- \otimes \mathcal{A}_0 \otimes \mathcal{A}_+, \mu_- \otimes \mu_0 \otimes \mu_+\big) \quad (46)$$

as input probability space of the channel $\kappa$, which satisfies $(A, \mathcal{A}, \mu) \in \mathfrak{P}(E_b)$. The resulting input-output probability space $(A \times Y, \mathcal{A} \otimes \mathcal{Y}, \mu\kappa)$ is well-defined since $\kappa$ is assumed to satisfy the measurability condition (M) w.r.t. $\mathcal{E}$. Let $\xi_0^b$ and $\eta_0^b$ denote the projections from $A \times Y$ onto $A_0 \subset X_0^b$ and $Y_0^b$, respectively, regarded as random variables on $(A \times Y, \mathcal{A} \otimes \mathcal{Y}, \mu\kappa)$. Further, let $\alpha$ and $\beta$ denote the functions defined on $A_0$ and $Y_0^b$, respectively, given by

$$\alpha(a_0) = i \quad \text{if} \quad a_0 = u_i \quad \text{and} \quad \beta(y_0^b) = j \quad \text{if} \quad y_0^b \in B_j.$$

(ii) *Applying Fano's inequality.* At first, we have

$$I\big(\alpha(\xi_0^b); \beta(\eta_0^b)\big) \leq I\big(\xi_0^b; \eta_0^b\big) \\ \leq C_b \\ \leq bC, \quad (47)$$

where the first inequality follows from the data processing inequality. Since $\mu \in \mathfrak{P}(E_b)$ belongs to the class of probability measures w.r.t. which $C_b$ is defined in Definition 2.19, we obtain the second inequality. We assume a stationary channel $\kappa$ and a family of input constraints $\mathcal{E}$ satisfying the regularity condition (R) in (8). Therefore, we can apply Lemma 2.21 and obtain the last inequality from (23).

If $p_e$ is given by

$$p_e = \mu\kappa\big(\alpha(\xi_0^b) \neq \beta(\eta_0^b)\big) \\ = \sum_{i=1}^m \mu\kappa\big(\alpha(\xi_0^b) = i, \beta(\eta_0^b) \neq i\big) \\ = \sum_{i=1}^m \mu\kappa\big([\{u_i\}] \times [B_i^c]\big), \quad (48)$$

then we obtain

$$h_m(p_e) \geq H\big(\alpha(\xi_0^b) | \beta(\eta_0^b)\big) \\ = \log m - I\big(\alpha(\xi_0^b); \beta(\eta_0^b)\big) \\ \geq \log m - bC \\ = \Big(1 - \frac{C}{\log(m)/b}\Big) \log m \\ \geq \Big(1 - \frac{C}{C + \rho}\Big) \log m. \quad (49)$$

Fano's inequality given in Lemma 2.26 yields the first inequality, where $h_m$ is defined in (27). The subsequent equality holds because we have $I\big(\alpha(\xi_0^b); \beta(\eta_0^b)\big) = H\big(\alpha(\xi_0^b)\big) + H\big(\alpha(\xi_0^b) | \beta(\eta_0^b)\big)$ and $\mu\kappa\big(\alpha(\xi_0^b) = i\big) = \mu\kappa\big(\xi_0^b = u_i\big) = \mu_0\big(\{u_i\}\big) = 1/m$. The next inequality follows from (47) and the last from the assumption $m \geq e^{(C+\rho)b}$.

Assume that we restrict the function $h_m$ to the interval $[0, (m-1)/m]$ and denote the corresponding inverse by $h_m^{-1}$. Then $h_m^{-1}$ is a well-defined, monotonically increasing function on $[0, \log m]$ with values in $[0, (m-1)/m]$. The following chain of inequalities holds

$$0 < h_{m-1}^{-1}\bigl(b\log(m-2)\bigr) \tag{50}$$
$$< h_m^{-1}\bigl(b\log(m-1)\bigr) \tag{51}$$
$$< h_m^{-1}\bigl(b\log m\bigr) \tag{52}$$

for all $0 < b \leq 1$ and $m \geq 4$. If $m = 2$, then the right-hand side of (50) and (51) have to be omitted and if $m = 3$, then only the right-hand side of (50) has to be omitted. Applying $h_m^{-1}$ to (49) and using (50)–(52) yields

$$p_e \geq h_{m_0}^{-1}\left(\left[1 - \frac{C}{C+\rho}\right]\log 2\right) > 0, \tag{53}$$

where $m_0 = 2$ if $m = 2$ and $m_0 = 3$ if $m > 2$. We have also used $(1 - C/(C+\rho)) \in (0,1)$.

(iii) *Evaluation of error probabilities.* We define for any $a_0 \in A_0$ and $B \in \mathcal{Y}_0^b$

$$\bar{\kappa}(a_0, B) = \int_{A_- \times A_+} \kappa\bigl((a_-, a_0, a_+), [B]\bigr)\,\mathrm{d}\mu_- \otimes \mu_+(a_-, a_+) \tag{54}$$

using the measures defined in (45) and the representation of an element $a \in A$ as 3-tuple $a = (a_-, a_0, a_+) \in A_- \times A_0 \times A_+$. The integral is well-defined since $\kappa$ is assumed to satisfy the measurability condition (M). Basic measure-theoretic considerations yield $\bar{\kappa}$ is a Markov kernel from $(A_0, \mathcal{A}_0)$ to $(Y_0^b, \mathcal{Y}_0^b)$. We have

$$\mu\kappa\bigl([\{u_i\}] \times [B_i^c]\bigr) = \int_{[\{u_i\}]} \kappa(a, [B_i^c])\,\mathrm{d}\mu(a)$$
$$= \int_{\{u_i\}} \bar{\kappa}(a_0, B_i^c)\,\mathrm{d}\mu_0(a_0)$$
$$= \frac{1}{m}\bar{\kappa}(u_i, B_i^c), \tag{55}$$

where the first equality is simply the definition of the channel input-output probability measure. From the specific product structure of $\mu$ defined in (46) we obtain the second equality using Fubini's theorem. The last equality is due to the definition of the measure $\mu_0$ in (44).

Inserting (55) into (48) and using (53) yields, that at least for one $i_0 \in \{1, 2, \ldots, m\}$ we have

$$\bar{\kappa}(u_{i_0}, B_{i_0}^c) \geq \epsilon^* > 0, \tag{56}$$

where $\epsilon^*$ is defined as the middle expression of (53). Using the product structure of $(A, \mathcal{A}, \mu)$ given in (46) and because $\kappa(\cdot, [B_{i_0}^c])$ is an $\mathcal{A}$-measurable function on $A$ due to condition (M) we can consider $\kappa\bigl((\cdot, u_{i_0}, \cdot), [B_{i_0}^c]\bigr)$ as random variable on the probability space $(A_- \times A_+, \mathcal{A}_- \otimes \mathcal{A}_+, \mu_- \otimes \mu_+)$. If the expectation of a nonnegative random variable is greater or equal to some constant, then the probability that the random variable has values greater or equal to this constant is positive. Since $\bar{\kappa}(u_{i_0}, B_{i_0}^c)$ is the expectation of $\kappa\bigl((\cdot, u_{i_0}, \cdot), [B_{i_0}^c]\bigr)$ we obtain from (56)

$$\mu_- \otimes \mu_+\Bigl(\kappa\bigl((\cdot, u_{i_0}, \cdot), [B_{i_0}^c]\bigr) \geq \epsilon^*\Bigr) > 0.$$

This implies, that at least for one $x \in U_b^* \cap [\{u_{i_0}\}]$ we have

$$\kappa(x, [B_{i_0}^c]) \geq \epsilon^* > 0,$$

where $U_b^* = A$ holds with $U_b^*$ as defined in (19). Thus we have $\varrho_{\max}(U_b^*) \geq \epsilon^*$ for the maximal decoding error, which completes the proof. ∎

**(3.3) Remark (Discussion, operational meaning of information rate capacity, average error probability).** Since we have $U_b^* \subset E_b^* \subset E'$ for the sets specified in (5) and (7) the weak converse holds if $U_b^*$ is replaced by $E_b^*$ or $E'$.

Clearly, the weak converse does only make sense for finite information rate capacity. It holds in particular under the conditions of Theorem 3.1, which also imply due to Lemma 2.8 that the channel $\kappa$ satisfies the required measurability condition (M) w.r.t. $\mathcal{E}$. The combination of the coding theorem and converse establishes the operational meaning of the information rate capacity as introduced in Definition 2.19. Indeed, from Theorem 3.1 we obtain

$$\overline{C} = \overline{C}(\mathcal{E}) = \inf_{\epsilon>0} \limsup_{b\to\infty} \frac{1}{b}\log M^*(b, E_b, E_b^*, \epsilon) \geq C,$$

i.e., the optimistic coding capacity $\overline{C}$ introduced in Paragraph 2.13 is lower-bounded by the information rate capacity $C$. Here we have additionally used the inequality $M^*(b, E_b, E', \epsilon) \leq M^*(b, E_b, E_b^*, \epsilon)$, which holds because $E_b^* \subset E'$. Further, from Theorem 3.2 and the comment at the beginning of this remark we obtain

$$\inf_{\epsilon>0} \sup_{b>0} \frac{1}{b}\log M^*(b, E_b, E_b^*, \epsilon) \leq C$$

and therefore

$$\overline{C} \leq C,$$

i.e., the optimistic coding capacity $\overline{C}$ is also upper-bounded by the information rate capacity $C$. Combining the coding theorem and the weak converse we therefore have

$$C = \overline{C}, \tag{57}$$

which establishes the operational meaning of $C$. Note, an optimal input distribution achieving $C$ is not required to exist within the class $\mathcal{P}(\mathcal{E})$ of finite alphabet block-memoryless sources satisfying the constraints $\mathcal{E}$. Therefore, weak continuity of the channel, separability of the input and output space, or the compactness of the input constraints as in [4] are not required to derive the coding theorem and converse.

Various important examples of channel models satisfying all the required properties in the coding theorem and converse are discussed in [5, Chs. IV and V] together with tools useful to study these properties.

Often, the pessimistic coding capacity – a term used in [58, p. 176] – is considered as alternative fundamental limit of reliable information transmission. In the setting of this paper it is defined by

$$\underline{C} = \underline{C}(\mathcal{E}) = \inf_{\epsilon>0} \liminf_{b\to\infty} \frac{1}{b}\log M^*(b, E_b, E_b^*, \epsilon).$$

In contrast to $\overline{C}$, it represents the supremum over all code rates $R$ such that for all $\epsilon > 0$ there exists a $b_0 \in T_+$ such that *for all $b \geq b_0$* there exists a $(b, E_b, E_b^*, \mathrm{e}^{Rb}, \epsilon)$-code. It follows that once a "good" code is found, then there exist good codes for *all* larger block length. In the general setting of this paper, it is open and beyond the scope of this work to characterize the pessimistic coding capacity in terms of an optimized measure of information. A seemingly obvious candidate would be to replace the information rate capacity by a quantity, where the limit superior in (22) is substituted by the limit inferior. Then, the quantity $s_0$ chosen in the beginning of the proof of Theorem 3.1 can be replaced by any $\tilde{s}_0 \geq s_0$. However, the probability measure $\mu$ chosen in the proof depends on $\tilde{s}_0$ and therefore also the quantity $n_0$ chosen later based on the ergodic theorem. Thus, with the approach at hand we cannot prove the "for all $b \geq b_0$"-part required to establish a pessimistic capacity (see Paragraph 2.13). In addition, the proof of the weak converse is based on the property, that limit superior and supremum are equal (see Lemma 2.21), which does not carry over to a situation, where we work with the limit inferior.

The coding theorem and converse are stated for the maximal decoding error probability $\varrho_{\max}(\cdot)$ as defined in (20). Alternatively, we can consider the average decoding error probability. Based on the notation of Definition 2.11 it is given by

$$\bar{\varrho}(\cdot) = \frac{1}{|\mathcal{C}|} \sum_{i=1}^{|\mathcal{C}|} \varrho(u_i, \cdot).$$

It follows immediately that Theorem 3.1 also holds for the average decoding error probability. To verify Theorem 3.2 for $\bar{\varrho}(U_b^*)$ we have to have a closer look. Let us follow the proof of the weak converse up to inequality (56). Then we observe that even

$$\frac{1}{m} \sum_{i=1}^{m} \bar{\kappa}(u_i, B_i^c) \geq \epsilon^* > 0$$

holds. Similar to the last part of the preceding proof we can therefore conclude

$$\bar{\varrho}(U_b^*) = \frac{1}{m} \sum_{i=1}^{m} \sup_{(a_-, a_+) \in A_- \times A_+} \kappa\big((a_-, u_i, a_+), [B_i^c]\big)$$
$$\geq \sup_{(a_-, a_+) \in A_- \times A_+} \frac{1}{m} \sum_{i=1}^{m} \kappa\big((a_-, u_i, a_+), [B_i^c]\big)$$
$$\geq \epsilon^* > 0.$$

This shows that Theorem 3.2 also holds for the average decoding error.

### III.B. On Measurability Requirements

As already addressed in Remark 2.2 Ahlswede pointed out in [2, Sec. 6.1] that for the model considered there – a discrete-time memoryless channel (see Remark 4.4) with abstract alphabet – the measurability of the channel w. r. t. the input signals is not necessary when studying coding problems. Let us explain this aspect and the difference to the situation in this paper, where a coding theorem and a converse are derived for channels, which are allowed to have also infinite input memory.

Generally speaking, if the channel $\kappa$ is memoryless, then we have to deal in all steps of the derivations only with input signals of finite duration and therefore only with a finite set $\{\kappa(a, \cdot)\}$ of probability measures defined by the channel. This set can always be considered as Markov kernel by using the power set of the relevant finite input signal set as $\sigma$-algebra. This way measurability issues do not arise. However, if the input memory is allowed to be infinite, then we have to deal with input signals of infinite duration so that an uncountable set $\{\kappa(a, \cdot)\}$ of probability measures has to be taken into account. Then measurability aspects come necessarily into play because utilizing the power set of an uncountable set of input signals as $\sigma$-algebra (w. r. t. which all functions would be measurable because it is the largest possible $\sigma$-algebra) does not work in this context.

Let us investigate one representative part of the proof of the coding theorem given in Theorem 3.1 to gain more inside into the preceding argumentation. In the proof we consider a probability measure $\mu_0$, that is concentrated on finitely many input signals in a certain time interval, i.e., $\mu_0$ has finite support. Looking into part (iv) of the proof we see that the probability measure $\mu'_0$ given in (39) has finite support as well, since it is defined on the finite space $(A'_0, \mathcal{A}'_0)$ as a product of shifted copies of $\mu_0$. If the channel $\kappa$ is memoryless, then $\kappa(\cdot, [B])$ is a function just on $(A'_0, \mathcal{A}'_0)$ for all $B \in \mathcal{Y}_0^{t_0}$, because for a memoryless channel the probability of events at the channel output in a certain time interval only depend on the channel input in the same time interval. In the remainder of the proof the relevant channel is therefore given by the finite set $\{\kappa(a'_0, \cdot), a'_0 \in A'_0\}$ of probability measures. In contrast, if the input memory is infinite, then in part (iv) of the proof the set $\{\kappa((a_-, a'_0), \cdot), a_- \in A_-, a'_0 \in A'_0\}$ of probability measures has to be taken into account, e.g., to define the induced channel $\bar{\kappa}$ in (40). The set of measures is uncountable because the set $A_-$ is uncountable (which might be surprising but indeed, the countable infinite product of finite sets is uncountable). Thus, for (40) to be well-defined we have to consider a measurability condition such as (M) defined in Paragraph 2.6. Under the requirements of the coding theorem regarding causality and type of input memory, however, the condition (M) is automatically satisfied so that it does not pose additional restrictions to derive the theorem. The previously described situation is similar when we look into part (iii) of the proof of the converse given in Theorem 3.2. Note that a causal channel with finite input memory behaves analog to a memoryless channel regarding the (non)requirement of a measurability condition.

We continue with a discussion on an alternative measurability condition considered in the literature and the drawbacks of this condition when studying coding problems. Recall that we consider a channel $\kappa$ with time structure as introduced in Definition 2.4 with input product space $X$ defined in (2) by the family $\{X_t, t \in T\}$ of sets with $X_t = X_0$ for all $t \in T$. Assume that we have given the family $\{\mathcal{X}_t, t \in T\}$ of $\sigma$-algebras, where $\mathcal{X}_t$ is a $\sigma$-algebra on $X_t$ with $\mathcal{X}_t = \mathcal{X}_0$ for all $t \in T$. We define the product-$\sigma$-algebra

$$\mathcal{X} = \bigotimes_{t \in T} \mathcal{X}_t,$$

and say the channel $\kappa$ satisfies the measurability condition

(M')    if for all $B \in \mathcal{Y}$ it holds that $\kappa(\cdot, B)$ as function on $X$ is $\mathcal{X}$-measurable.

Condition (M′) is in certain situations the natural choice because the product $\sigma$-algebra $\mathcal{X}$ is the typical $\sigma$-algebra for which the distribution of a random process with time index set $T$ is defined. For example, if the result in [5, Thm. 13.11] is used to study the memory properties of the pair random process at the channel input and output, then (M′) is the adequate condition.

As shown in [5] the derivations of the coding theorem and converse also work if the channel $\kappa$ satisfies the condition (M′). However, using condition (M′) outer measures and a special form of Feinstein's lemma have to be used in [5], which is not required in this paper, leading to simplified and more transparent derivations. In addition, for (M′) there is no result such as Lemma 2.8, which would guarantee the validity of (M′) if the channel is asymptotically input-memoryless (and causal) – a property required anyway to proof the coding theorem of this paper. Furthermore and most importantly, there are relevant continuous-time channel models, that do not satisfy (M′) but satisfy (M). Consider the following example.

**(3.4) Example (Continuous-time linear filter satisfying (M) but not (M′)).** Let $\kappa$ be a continuous-time channel with real-valued input and output signals, i.e., in Definition 2.4 we fix $T = \mathbb{R}$ and $X_0 = Y_0 = \mathbb{R}$, where $Y_0$ is equipped with the corresponding Borel-$\sigma$-algebra. Further, let $h : [0, \infty) \to \mathbb{R}$ be a Lebesgue-measurable, absolutely integrable function and assume that the channel output $y = \{y_t, t \in \mathbb{R}\}$ is defined by

$$y_t = \int_0^\infty h_\tau x_{t-\tau} \, d\tau, \qquad t \in \mathbb{R}, \tag{58}$$

for all input signals $x = \{x_t, t \in \mathbb{R}\} \in X$, for which the integrals in (58) exist. Otherwise, the output is arbitrarily defined, e.g., equal to some constant function. This deterministic special case of a channel represents a continuous-time linear time-invariant filter with impulse response $h$. The channel is causal, stationary and output-memoryless. If the impulse response has finite support, then the channel has finite input memory. If the impulse response has infinite duration, then further conditions are necessary for the channel to be asymptotically input-memoryless, e.g., if we additionally have additive noise and suitable constraints on the channel input space.

Consider now the family $\mathcal{E} = \{E_s, s \in T_+\}$ of input constraints, where $E_s$ is given by

$$E_s = \big\{ x \in X_0^s : x \text{ is Lebesgue-measurable} \\ \text{and} \sup_{t \in (0,s]} |x_t| \leq c \big\},$$

with c being some positive constant, i.e., we restrict the channel input to Lebesgue-measurable signals bounded in amplitude by c. It is not difficult to verify, that $\mathcal{E}$ satisfies the regularity condition (R) in (8). Furthermore, it is not difficult to see that $\kappa$ satisfies the measurability condition (M) w.r.t. $\mathcal{E}$, which is particularly the case because the $\sigma$-algebra at the channel output is a product-$\sigma$-algebra as introduced in (3). Indeed, it is sufficient to show that $\kappa(\cdot, B)$ as function on $A$ is $\mathcal{A}$-measurable for all $(A, \mathcal{A}) \in \mathfrak{A}(\mathcal{E})$ and $B \in \mathfrak{C}'_Y$ with $\mathfrak{A}(\mathcal{E})$ and $\mathfrak{C}'_Y$ defined in (13) and (14). Therefore, it is sufficient to show that the integral operator defined by (58) for fixed $t \in \mathbb{R}$ is measurable w.r.t. to all input spaces $(A, \mathcal{A}) \in \mathfrak{A}(\mathcal{E})$. This is shown similar to the proof of Lemma 2.8 given in Paragraph A.1 using the boundedness of the inputs and the absolute integrability of the impulse response. In contrast, if the input space $X$ is equipped with the product-$\sigma$-algebra $\mathcal{X} = \mathcal{Y}$, then – as a matter of principle – $\kappa$ is not measurable w.r.t. $\mathcal{X}$ (see [66, Sec. 5.1]), i.e., $\kappa$ does not satisfy the condition (M′). This is true even if the impulse response has finite support, i.e., if the channel has finite input memory.

The coding results of Kadota and Wyner [3] for continuous-time channels with real-valued input and output signals are derived under the assumption that (M′) holds, where $\mathcal{X}_t$ is taken as the usual Borel-$\sigma$-algebra on $X_t = \mathbb{R}$. However, a continuous-time linear filter is used in [3] to motivate the analysis in the paper, which is inconsistent with the type of measurability assumed there since this filter does not satisfy (M′) as discussed in the previous Example 3.4. Being able to use condition (M) in the present paper is closely related to the way we introduce the information rate capacity. Based on the definition of information rate capacity in [3] condition (M) cannot be used.

## IV. OUTPUT MEMORY CONDITIONS

Output memory conditions characterize – for given channel input – the (asymptotic) independence of sufficiently time-separated channel output events. There is a great variety of such conditions that are either based on ergodic-theoretic mixing properties or on so-called strong mixing properties specified by stochastic dependence coefficients. In [5, § 13] such conditions are analyzed and reviewed in detail. By the hierarchy derived in [5, Thm. 13.10] it is shown how these conditions are related and that they all imply the ergodicity of a channel with time structure as required in Theorem 3.1. In particular, it is shown how much weaker the ergodicity condition (2.7.ii) is, compared to the output memory condition considered in the coding theorem in [3], called $\psi$-mixing condition. The ergodicity condition (2.7.ii) is the least restrictive condition possible to apply Pinsker's ergodic theorem of information theory in the present context. In contrast, the $\psi$-mixing condition, which is defined using the $\psi$-dependence coefficient, is significantly more restrictive as will be demonstrated below by showing that for the important class of Gaussian channels it is in fact equivalent to finite output memory. Based on the example of a simple, practically relevant additive Gaussian noise channel with rational spectral density we show that the coding theorem in [3] does not apply to important channel models, for which Theorem 3.1 still holds.

As before $\kappa$ is a channel with time structure as introduced in Definition 2.4 with input product space $X$ and output product measurable space $(Y, \mathcal{Y})$. Further, $\eta = \{\eta_t, t \in T\}$ denotes the family of coordinate projections on the channel output space, where $\eta_t$ denotes the projection from $Y$ onto $Y_t$, and $\eta_u^v$ denotes the projection from $Y$ onto $Y_u^v$ for all $u < v \in \overline{T}$. Whenever $Y_t$ is equal to the set of real numbers or vectors, then we associate with it the corresponding Borel-$\sigma$-algebra.

**(4.1) Definition ($\psi$-dependence coefficient).** Let $\alpha$ and $\beta$ be random variables on the probability space $(\Omega, \mathcal{F}, \mathrm{P})$ with values in the measurable spaces $(\Omega', \mathcal{F}')$ and $(\Omega'', \mathcal{F}'')$, respectively. Then the $\psi$-dependence coefficient of $\alpha$ and $\beta$ is defined by

$$\psi(\alpha; \beta) = \sup \left| \frac{\mathrm{P}_{\alpha,\beta}(F \times G)}{\mathrm{P}_\alpha(F) \mathrm{P}_\beta(G)} - 1 \right|,$$

where the supremum is taken w.r.t. all $F \in \mathcal{F}'$ and $G \in \mathcal{F}''$ satisfying the condition $\mathrm{P}_\alpha(F)\mathrm{P}_\beta(G) > 0$. By $\mathrm{P}_{\alpha,\beta}$, $\mathrm{P}_\alpha$, and $\mathrm{P}_\beta$ we denote the distribution of $(\alpha, \beta)$, $\alpha$, and $\beta$, respectively.

**(4.2) Remark.** Recall that the mutual information measures the dependence between two random variables since it measures the difference between the joint distribution and the product of the marginal distributions in terms of relative entropy. The $\psi$-dependence coefficient quantifies this difference in another way. It represents therefore an alternative measure of stochastic (in)dependence. It is used below to define the channel output memory condition considered in [3].

**(4.3) Definition (Finite output memory, $\psi$-mixing channel).** The channel $\kappa$ has finite output memory if for all $s \in T$ there exists a $t_o(s) \in T_0$ such that for any $B \in \mathcal{Y}_-^s$, $\hat{B} \in \mathcal{Y}_{s+t_o(s)}^+$, and $x \in X$ we have

$$\kappa(x, [B] \cap [\hat{B}]) = \kappa(x, [B]) \kappa(x, [\hat{B}]).$$

For given $s \in T$ we call the smallest possible $t_o(s)$ the output memory length at time $s$. If $t_o(s) = 0$ for all $s \in T$, then we say the channel is output-memoryless.

The channel $\kappa$ is called $\psi$-mixing if for all $s \in T$

$$\lim_{t \to \infty} \sup_{x \in X} \psi\bigl(\eta_-^s; \eta_{s+t}^+ \,\big|\, x\bigr) = 0,$$

where $\psi\bigl(\eta_-^s; \eta_{s+t}^+ \,\big|\, x\bigr)$ denotes the $\psi$-dependence coefficient of $\eta_-^s$ and $\eta_{s+t}^+$, considered as random variables on the channel output probability space $(Y, \mathcal{Y}, \kappa(x, \cdot))$.

**(4.4) Remark (Alternative definition, interpretation, memoryless channels).** Note that we have the following equivalent definition of finite output memory: The channel $\kappa$ has finite output memory if for all $s \in T$ there exists a $t_o(s) \in T_0$ such that for all $x \in X$ the random variables $\eta_-^s$ and $\eta_{s+t_o(s)}^+$ are independent. Finite channel output memory means, sufficiently time-separated output events are independent given a fixed input. The $\psi$-mixing condition, in turn, means that for given input, future and present outputs are asymptotically independent from outputs remote in the past, where the (in)dependence is measured in terms of the $\psi$-dependence coefficient. In [3] the $\psi$-mixing channels are introduced under the name asymptotically output-memoryless channels. We have chosen a reformulation of this property in terms of the $\psi$-dependence coefficient because this allows us to exploit more easily the connections to the rich field of strong mixing conditions [67].

In contrast to the generality of the model considered in this paper, the most widely used model in the information-theoretic literature is a discrete-time, causal, stationary, input-memoryless (see Definition 2.7) and output-memoryless channel, often with finite alphabets.

**(4.5) Definition (Gaussian channel).** Suppose we have real (vector-) valued output signals, i.e., $Y_0 = \mathbb{R}^n$ for some positive integer $n$. If for all input signals $x \in X$ the family $\eta = \{\eta_t, t \in T\}$ of coordinate projections defined on the channel output probability space $(Y, \mathcal{Y}, \kappa(x, \cdot))$ is a Gaussian (vector) process, then $\kappa$ is called Gaussian channel.

**(4.6) Remark.** For the definition of a Gaussian random process see [5, Appendix A.6] and the references given there. We assume that all second moments of the involved random processes are finite, which is convenient for analysis and does not pose restrictions to models of practical situations. We restrict ourselves to the real case, however, the extension to the complex case is canonical. Note that we have not specified the input alphabet in the definition. The additive Gaussian noise channel considered in Example 4.8 below with $X_0 = Y_0 = \mathbb{R}$ is the most prominent example of a Gaussian channel. However, the class of Gaussian channels is much larger, including for example Gaussian (Rayleigh) fading channels.

The next theorem demonstrates how restrictive the output memory condition of Kadota and Wyner [3] is. Actually, it shows that for the important class of Gaussian channels the condition is equivalent to finite output memory. The result was derived by the authors in [68] and the proof is given in Paragraph B.2 of Appendix B. Further, with Example 4.8 we provide a simple, practically relevant channel with infinite output memory, which is excluded by the coding theorem in [3], however, for which Theorem 3.1 holds.

**(4.7) Theorem ($\psi$-mixing Gaussian channels have finite output memory).** Let $\kappa$ be a Gaussian channel. Then $\kappa$ is $\psi$-mixing if and only if $\kappa$ has finite output memory.

**(4.8) Example (Additive Gaussian noise).** Assume that the channel $\kappa$ has real-valued input and output signals, i.e., $X_0 = Y_0 = \mathbb{R}$. Let $(Z, \mathcal{Z})$ denote the product measurable space generated by the family $\{(Z_t, \mathcal{Z}_t), t \in T\}$ of measurable spaces, for which $(Z_t, \mathcal{Z}_t) = (Z_0, \mathcal{Z}_0)$, where $Z_0 = \mathbb{R}$ with $\mathcal{Z}_0$ being the usual Borel-$\sigma$-algebra on $\mathbb{R}$. Further, let $\zeta = \{\zeta_t, t \in T\}$ denote the family of coordinate projections defined on $Z$, where $\zeta_t$ denotes the projection from $Z$ onto $Z_t$. Assume that $\lambda$ is a probability measure on $(Z, \mathcal{Z})$.

Suppose the family $\eta = \{\eta_t, t \in T\}$ of coordinate projections on the channel output space is given for any $x \in X$ by

$$\eta = x + \zeta,$$

where addition is defined pointwise, i.e., $\eta_t = x_t + \zeta_t$ for all $t \in T$. That means $\kappa(x, \cdot) = \lambda_{x+\zeta}$ is equal to the distribution of the random process $x + \zeta$ for all $x \in X$. This channel is an additive noise channel.

Assume that $\zeta$ is a stationary zero mean, second order Gaussian process. Then $\eta$ is a Gaussian process with mean $x$ and the same covariance function as $\zeta$. If the support of the covariance function of the additive noise $\zeta$ is not a finite interval, then $\kappa$ does not satisfy the finite output memory condition as can be seen from the equivalent definition given in Remark 4.4. Thus, $\kappa$ is not $\psi$-mixing according to Theorem 4.7. As an example assume that $\zeta$ has a rational spectral density which is truly rational, i.e., which has an autoregressive part. The most simple process of this type is an autoregressive process of order one as given, e.g., in [5, Exm. 12.4.ii and Exm. C.4]. Then we have the previously described situation, which means for such channels the version of the coding theorem given in [3] is not applicable. However, Theorem 3.1 is applicable because the required ergodicity condition and also the remaining conditions on the channel are satisfied as shown in [5, Par. 16.3], where this example is studied in detail. Stationary Gaussian noise with a rational spectral density results from passing stationary white Gaussian noise through a (well-behaved) linear filter. It is therefore relevant to model various practical situations. This example demonstrates that the original formulation of the coding theorem in [3] for $\psi$-mixing channels does not apply to important channel models, which are covered by Theorem 3.1.

## V. Discussion and Summary of Contributions

Essentially, the main contribution of this paper is a generalization of the work of Kadota and Wyner [3], formulated in rigorous mathematical terms. The theorems in Section III generalize the work in [3] in a number of respects, namely in terms of channel model, input constraints, definition of information rate capacity, and, most importantly, w.r.t. required channel properties. Let us summarize the relevant extensions and modifications in relation to this work. In [3] a continuous-time channel with real-valued input and output signals is considered and a coding theorem and converse for those channels is formulated. In contrast, the theorems in Section III apply to channels with time structure in general, i.e., to discrete- as well as continuous-time channels with completely arbitrary alphabets. A consequent measure-theoretic description, in particular the use of general product spaces allows this generalized and unified formulation. Further restrictions of the alphabets, e.g., to separable metric spaces as in [4], are not required.

For Theorem 3.1 as well as the coding theorem in [3] the considered channel is required to be stationary, causal, and asymptotically input-memoryless. Clearly, finite input memory is included as special case since it implies asymptotic input-memorylessness, which is in general a form of infinite input memory. In addition, both theorems include a condition concerning the channel output memory, however, of fairly different quality. A property called asymptotic output-memorylessness is used in [3], which is called $\psi$-mixing condition in Definition 4.3. This output memory condition is quite restrictive, in particular for the important class of Gaussian channels. In Theorem 4.7 we show that for Gaussian channels it is indeed equivalent to finite output memory. As a result, the version of the theorem in [3] is for example not applicable to the important stationary additive Gaussian noise channel with rational noise spectral density as discussed in Example 4.8.

The condition of total ergodicity w.r.t. $\mathfrak{P}(\mathcal{E})$, i.e., w.r.t. all finite alphabet block-memoryless input sources satisfying the constraints $\mathcal{E}$, used in Theorem 3.1 is much weaker than the $\psi$-mixing condition as demonstrated by the hierarchy in [5, Thm. 13.9]. It is also weaker than the condition used in [5, Def. 2.7] since only finite alphabet sources are taken into account. As demonstrated in [5, Par. 16.4] it is also strictly weaker than the total ergodicity considered in [69, Def. 1] or [24, p. 360]. We are able to prove the coding theorem under the weaker condition by introducing a suitable definition of information rate capacity in Definition 2.19. Indeed, this allows us to apply a quite simple version of the ergodic theorem of information theory as given in Theorem 2.24 and to use identical derivations for finite and infinite capacity. Both is not possible with the characterization of information rate capacity considered in [3], where the more restrictive $\psi$-mixing condition is required to achieve that some version of the ergodic theorem can be applied. Details on the different alternatives of information rate capacity and the impact on proving the coding theorem are studied in [5, §10]. See also the discussion in [5, p. 57] on the requirement to modify the capacity definition such that the total ergodicity property can be employed and that simple ergodicity is not sufficient, as claimed in [12, p. 296], where discrete-time finite alphabet channels are considered.

Further improvements are based as well on the restriction in Definition 2.19 to finite alphabet block-memoryless input sources. On the one hand, we can employ the measurability condition (M) instead of (M'), which is a significant advantage because (M) can be verified – in contrast to (M') required by Kadota and Wyner [3] – for important continuous-time channel models such as linear filters. Additionally, (M) is implied by a channel input memory condition required anyway to prove the coding theorem, i.e., it does not impose an additional constraint. Using (M) instead of (M') and incorporating all associated modifications – including the avoidance of outer measures – represents essentially the evolution of the results in this paper compared to those given in the Ph.D. dissertation [5, Ch. II]. On the other hand, derivations become more transparent and simple because we can omit the use of so-called standard extensions of the input space as required in [3] to incorporate input constraints. Definition 2.19 is closely related to its operational meaning as optimistic coding capacity (see Remark 3.3). Using this specific form is inspired by the work in [70]. For the special case of a discrete-time memoryless channel also in [34, p. 318/324], [16] or [71] an information capacity definition based on finitely supported input measures is considered. Note that in the general setting of this paper, it is open to characterize the pessimistic coding capacity in terms of a suitable information capacity quantity.

The weak converse in Theorem 3.2 applies to all stationary channels with time structure. No further channel properties are required. In [3] no differentiated formulations of the coding theorem and converse regarding channel properties are given. In [12, p. 303] it is already noted that the converse in [3] should hold for all stationary channels.

Due to Definition 2.11 we have a convenient flexibility in taking the robustness of a block code w.r.t. past input signals into account and input constraints are incorporated in Theorem 3.1 and Theorem 3.2 in an abstract form. In order to prove the theorems the constraints have to satisfy the regularity condition (R) specified in (8). In [3] input constraints are characterized based on functionals as specified in the last part of Paragraph 2.5 but no regularity condition is mentioned in [3], although the considered examples satisfy the required condition. Additional requirements for the input constraints such as compactness like in [4] are not required. Further note that also the monotonicity result of Lemma 2.16 is employed in [3], but the proof in [3, Appendix II] is incorrect as demonstrated with the counterexample in Appendix C. See Paragraph A.2 for a rigorous proof of the statement.

We remark that in [3] also incremental versions of the channel properties are considered by using a $\sigma$-algebra of increments at the channel output. The motivation behind this extension is a mathematically rigorous treatment of continuous-time additive white noise. For the special case of real-valued output signals we can immediately apply this modification to the model used in this paper, without any change in the proof of the coding theorem and converse.

Most literature on coding theorems for continuous-time channels is restricted to additive noise or additive Gaussian noise. In contrast, the theorems in this paper apply to general continuous-time channel models. Based on a consequent measure-theoretic description and following the path in [3] we are able to avoid the commonly used transformation of continuous-time channels proposed by Holsinger [33] and Gallager [34, Ch. 8]. As a result, we can treat continuous-time channels with the same methods applicable to discrete-time channels. It is an inherent part of our model to take interference between successive codewords into account, which is an additional advantage compared to the approach in [34, Ch. 8]. Furthermore, we incorporate input constraints in a general and flexible form. These are important arguments in favor of the path taken in this paper.

The channel properties we require to prove the coding theorem characterize a large class of practically relevant communication models. Indeed, causality is always physically justified in the context of transmission over time. If a channel code is to be used repeatedly, stationarity is a typical assumption. However, studying coding problems for non-stationary channels within the general framework of this paper would be of significant practical relevance as well. Regarding channel output memory the considered condition of total ergodicity w.r.t. $\mathfrak{P}(\mathcal{E})$ is rather weak. Verifying the asymptotic input-memorylessness for specific channels, however, can be a very challenging task such that it is definitely worth to investigate more relaxed conditions for the channel input memory when studying coding problems. A generalization of the $\bar{d}$-continuity condition considered in [12] for discrete-time finite alphabet channels might be a possible approach. Various important examples of channel models satisfying all the mentioned properties required in Theorem 3.1 are discussed in [5, Chs. IV and V] together with tools useful to study these properties.

Finally note, the $\sigma$-algebra considered at the channel output determines the class of possible decoding sets taken into account for the analysis. In this paper we have chosen for the channel with time-structure the product-$\sigma$-algebra $\mathcal{Y} = \bigotimes_{t \in T} \mathcal{Y}_t$ (see (3) and Definition 2.4) since it is the natural $\sigma$-algebra for random sequences and processes. Thus, the decoding sets (or rather the corresponding cylinder sets) are elements of the product-$\sigma$-algebra $\mathcal{Y}$. In the continuous-time case it might be desirable to consider decoding sets from a larger $\sigma$-algebra, which is, e.g., a separable or measurable standard extension of the product-$\sigma$-algebra $\mathcal{Y}$. Therefore, an extension of the results of this paper to an enlarged output $\sigma$-algebra is also of interest for future work.

## APPENDIX

### APPENDIX A. PROOFS OF LEMMAS

**(A.1) Proof of Lemma 2.8.** Due to the comment below the definition of condition (M) in Paragraph 2.6 it is sufficient to show that $\kappa(\cdot, G)$ as function on $A$ is $\mathcal{A}$-measurable for all $(A, \mathcal{A}) \in \mathfrak{A}(\mathcal{E})$ and $G \in \mathfrak{C}_Y$, where $\mathfrak{A}(\mathcal{E})$ and $\mathfrak{C}_Y$ are defined in (13) and (14).

Let $G \in \mathfrak{C}_Y$ be arbitrary. Then there exist $u < v \in T$ such that $G \in [\mathcal{Y}_u^v]$. Further, let $(A, \mathcal{A}) \in \mathfrak{A}(\mathcal{E})$ be an arbitrary product measurable space as defined in (13), where $s \in T_+$ is the time duration of the signals in $A_0$. Then there exist $i < j \in \mathbb{Z}$ with $is < u$ and $js > v$ such that $G \in [\mathcal{Y}_{is}^{js}]$. We consider $\kappa(\cdot, G)$ as function on $A \subset X$ and fix an arbitrary signal $a = \{a_k, k \in \mathbb{Z}\} \in \times_{k \in \mathbb{Z}} \langle A_0 \rangle_{ks} = A$. For any $n \in \mathbb{N}$ we define the function $g_n$ on $A$ by

$$g_n(x) = \kappa\big((a_-^{i-n}, x_{i-n}^+), G\big),$$

for all $x = \{x_k, k \in \mathbb{Z}\} \in A$, where $a_-^{i-n} = \{a_k, k \leq i-n\}$ denotes the projection of $a$ onto $A_-^{i-n}$ and $x_{i-n}^+ = \{x_k, k > i-n\}$ denotes the projection of $x$ onto $A_{i-n}^+$.

Due to the definition of $g_n$ and the causality of $\kappa$ in combination with $G \in [\mathcal{Y}_{is}^{js}]$ we have that $g_n(x)$ does only depend on coordinates of $x$ with index $i - n < k < j$. In addition, the product $\sigma$-algebra $\mathcal{A}_{i-n}^j$ is equal to the power set $2^{A_{i-n}^j}$, since $A_0$ is a finite set. Thus, $g_n$ is $[\mathcal{A}_{i-n}^j]$-measurable and therefore $\mathcal{A}$-measurable. Since $\kappa$ is asymptotically input-memoryless for the signal set $E'$ we further have

$$g_n \xrightarrow[(n \to \infty)]{} \kappa(\cdot, G)$$

pointwise for all $x \in A$. As a standard result from measure theory we obtain from this convergence and the $\mathcal{A}$-measurability of $g_n$ for all $n \in \mathbb{N}$ the $\mathcal{A}$-measurability of $\kappa(\cdot, G)$. As $G \in \mathfrak{C}_Y$ and $(A, \mathcal{A}) \in \mathfrak{A}(\mathcal{E})$ were chosen arbitrary, we obtain that $\kappa$ satisfies the condition (M) w.r.t. $\mathcal{E}$.

**(A.2) Proof of Lemma 2.16.** In the subsequent proof, if $m > n$ and $\xi_m^n$ is the condition for a conditional mutual information, then this denotes the corresponding unconditional mutual information, for example $I(\eta_1; \eta_2 | \xi_1^0) = I(\eta_1; \eta_2)$.

(i) *Proof of (2.16.i).* First assume that $I(\xi_0^n; \eta_0^n) < \infty$ for all $n \in \mathbb{N}$. To show that the sequence $\{n^{-1} I(\xi_0^n; \eta_0^n), n \in \mathbb{N}\}$ is monotonically increasing, we show that

$$\frac{1}{n+1} I(\xi_0^{n+1}; \eta_0^{n+1}) - \frac{1}{n} I(\xi_0^n; \eta_0^n) \geq 0 \quad (59)$$

holds for all $n \in \mathbb{N}$. Assume that we have the identity

$$n\, I(\xi_0^{n+1}; \eta_0^{n+1}) - (n+1)\, I(\xi_0^n; \eta_0^n) = \sum_{k=1}^{n} (n-k+1)\, I(\xi_k; \eta_{n+1} | \eta_0^n, \xi_0^{k-1}) \quad (60)$$
$$+ \sum_{k=2}^{n+1} (k-1)\, I(\xi_k; \xi_1, \eta_1 | \eta_1^{n+1}, \xi_1^{k-1}).$$

Then, dividing (60) by $n(n+1)$ and using the nonnegativity of the conditional mutual information yields (59). To obtain (60) we first apply the chain rules of (conditional) mutual information repeatedly to obtain

$$n\, I(\xi_0^{n+1}; \eta_0^{n+1}) \quad - (n+1)\, I(\xi_0^n; \eta_0^n)$$
$$= n \sum_{k=1}^{n+1} I(\xi_k; \eta_0^{n+1} | \xi_0^{k-1}) - (n+1) \sum_{k=1}^{n} I(\xi_k; \eta_0^n | \xi_0^{k-1}). \quad (61)$$

We can rewrite (61) as

$$\sum_{k=1}^{n} (n-k+1) \Big[ I(\xi_k; \eta_0^{n+1} | \xi_0^{k-1}) - I(\xi_k; \eta_0^n | \xi_0^{k-1}) \Big]$$
$$+ \sum_{k=2}^{n+1} (k-1) \Big[ I(\xi_k; \eta_0^{n+1} | \xi_0^{k-1}) - I(\xi_{k-1}; \eta_0^n | \xi_0^{k-2}) \Big], \quad (62)$$

which can be seen by expanding the sums in (61) as shown in Table I. Applying the chain rules of (conditional) mutual information to the rows in Table I marked by $(*)$ yields

$$(n-k+1) \Big[ I(\xi_k; \eta_0^{n+1} | \xi_0^{k-1}) - I(\xi_k; \eta_0^n | \xi_0^{k-1}) \Big]$$
$$= (n-k+1)\, I(\xi_k; \eta_{n+1} | \eta_0^n, \xi_0^{k-1}) \quad (63)$$

for $k = 1, 2, \ldots, n$. Applying the chain rules to the rows in Table I marked by $(\circ)$ yields

$$(k-1)\Big[I(\xi_k; \eta_0^{n+1} | \xi_0^{k-1}) \quad - I(\xi_{k-1}; \eta_0^n | \xi_0^{k-2})\Big]$$
$$= (k-1)\Big[I(\xi_k; \xi_1, \eta_0^{n+1} | \xi_1^{k-1}) \quad - I(\xi_k; \xi_1 | \xi_1^{k-1})$$
$$\qquad - I(\xi_k; \eta_1^{n+1} | \xi_1^{k-1})\Big]$$
$$= (k-1)\Big[I(\xi_k; \xi_1, \eta_1 | \eta_1^{n+1}, \xi_1^{k-1}) - I(\xi_k; \xi_1 | \xi_1^{k-1})\Big]$$
$$= (k-1)\, I(\xi_k; \xi_1, \eta_1 | \eta_1^{n+1}, \xi_1^{k-1}) \quad (64)$$

for $k = 2, 3, \ldots, n+1$. For the first equality we have additionally used

$$I(\xi_{k-1}; \eta_0^n | \xi_0^{k-2}) = I(\xi_k; \eta_1^{n+1} | \xi_1^{k-1}),$$

which follows from the stationarity of the pair sequence $\{(\xi_k, \eta_k), k \in \mathbb{Z}\}$. For (64) we have used

$$I(\xi_k; \xi_1 | \xi_1^{k-1}) = I(\xi_1^k; \xi_1) - I(\xi_1^{k-1}; \xi_1) = 0,$$

where the first equality follows again from the chain rules. The second equality holds due to the condition in the lemma that $\xi = \{\xi_k, k \in \mathbb{N}\}$ is an independent sequence of random variables such that the involved mutual information quantities are zero. Now combining (61)–(64) yields the identity in (60). The shown monotonicity clearly implies

$$\bar{I}(\xi; \eta) = \lim_{n \to \infty} \frac{1}{n} I(\xi_0^n; \eta_0^n) = \sup_{n \geq 1} \frac{1}{n} I(\xi_0^n; \eta_0^n). \quad (65)$$

The above derivations do not contain any indeterminate expression since from the initial assumption of the proof we have $I(\xi_0^{n+1}; \eta_0^{n+1}) < \infty$, which implies the finiteness of all information quantities considered above. Now let us consider the case when the assumption $I(\xi_0^n; \eta_0^n) < \infty$ is not true for all $n \in \mathbb{N}$. Then, because the sequence $\{I(\xi_0^n; \eta_0^n), n \in \mathbb{N}\}$ is monotonically increasing in $n$, there exists an $n_0 \in \mathbb{N}$ such that $I(\xi_0^n; \eta_0^n) < \infty$ for all $n < n_0$ and $I(\xi_0^n; \eta_0^n) = \infty$ for all $n \geq n_0$. In this situation inequality (59) is valid for all $n < n_0$, because for all $n < n_0 - 1$ we have the same situation as before and for $n = n_0 - 1$ we have $\infty \geq 0$. If $n \geq n_0$, then we have $n^{-1} I(\xi_0^n; \eta_0^n) = \infty$. (If $n_0 = 1$ or $n_0 = 2$ the previous discussion simplifies accordingly.) Thus, we have the monotonicity of the sequence $\{n^{-1} I(\xi_0^n; \eta_0^n), n \in \mathbb{N}\}$. Finally, (65) still holds since both sides are infinite. This completes the proof of the first part of Lemma 2.16.

(ii) *Proof of (2.16.ii).* For $m, n \in \mathbb{N}$ we have

$$I(\xi_0^{m+n}; \eta_0^{m+n}) = I(\xi_0^m, \xi_m^{m+n}; \eta_0^m, \eta_m^{m+n})$$
$$\geq I(\xi_0^m; \eta_0^m) + I(\xi_m^{m+n}; \eta_m^{m+n})$$
$$= I(\xi_0^m; \eta_0^m) + I(\xi_0^n; \eta_0^n).$$

Applying the nonnegativity of conditional mutual information and the chain rules in combination with the condition in the lemma that $\xi = \{\xi_k, k \in \mathbb{N}\}$ is an independent sequence of random variables (see [72, Lem. 4.2.2]) yields the inequality. The last equality follows from the stationarity of the pair sequence $\{(\xi_k, \eta_k), k \in \mathbb{Z}\}$. Thus, the sequence $\{I(\xi_0^n; \eta_0^n), n \in \mathbb{N}\}$ is superadditive. If $I(\xi_0^n; \eta_0^n) < \infty$ for all $n \in \mathbb{N}$, then

$$\bar{I}(\xi; \eta) = \lim_{n \to \infty} \frac{1}{n} I(\xi_0^n; \eta_0^n) = \sup_{n \geq 1} \frac{1}{n} I(\xi_0^n; \eta_0^n) \quad (66)$$

follows from Fekete's lemma given, e.g., in [73, p. 3] Otherwise, both sides of (66) are infinite. (See the discussion at the end of part (i).) This completes the proof of the second part of Lemma 2.16.

**(A.3) Proof of Lemma 2.21.** (i) *Proof of (23).* Let us define $\gamma = \sup_{s \in T_+} C_s / s$ and let $\epsilon > 0$. If $\gamma < \infty$ we put $\rho = \gamma - \epsilon$ and if $\gamma = \infty$ let $\rho > 0$ be arbitrary. Then due to the definition of $\gamma$ and $C_s$ in Definition 2.19 there exists in both cases an $s_0 \in T_+$ and a channel input probability space $(A, \mathcal{A}, \mu) \in \mathfrak{P}(E_{s_0})$ such that

$$\rho \leq \frac{1}{s_0} I\big(\xi_0^{s_0}; \eta_0^{s_0}\big). \quad (67)$$

For all $k \in \mathbb{Z}$ we denote by

$$\alpha_{k+1} = \xi_{ks_0}^{(k+1)s_0} \qquad \text{and} \qquad \beta_{k+1} = \eta_{ks_0}^{(k+1)s_0} \quad (68)$$

the projections from $A \times Y$ onto $A_k = \langle A_0 \rangle_{ks_0}$ and $Y_{ks}^{(k+1)s}$, respectively. Then $\alpha = \{\alpha_k, k \in \mathbb{Z}\}$ and $\beta = \{\beta_k, k \in \mathbb{Z}\}$ are random sequences on the channel input-output probability space $(A\times Y, \mathcal{A}\otimes\mathcal{Y}, \mu\kappa)$, where the probability measure $\mu\kappa$ is well-defined since we assume that $\kappa$ satisfies the measurability condition (M) w.r.t. $\mathcal{E}$.

| $k$ | | | | |
|---|---|---|---|---|
| 1 | $n\,I(\xi_1;\eta_0^{n+1})$ | $-$ | $n\,I(\xi_1;\eta_0^n)$ | $(*)$ |
| 2 | $+\quad I(\xi_2;\eta_0^{n+1}|\xi_1)$ | $-$ | $I(\xi_1;\eta_0^n)$ | $(\circ)$ |
|   | $+(n-1)\,I(\xi_2;\eta_0^{n+1}|\xi_1)$ | $-(n-1)\,I(\xi_2;\eta_0^n|\xi_1)$ | | $(*)$ |
| 3 | $+\quad 2\,I(\xi_3;\eta_0^{n+1}|\xi_0^2)$ | $-$ | $2\,I(\xi_2;\eta_0^n|\xi_1)$ | $(\circ)$ |
|   | $+(n-2)\,I(\xi_3;\eta_0^{n+1}|\xi_0^2)$ | $-(n-2)\,I(\xi_3;\eta_0^n|\xi_0^2)$ | | $(*)$ |
|   |   | $-$ | $3\,I(\xi_3;\eta_0^n|\xi_0^2)$ | $(\circ)$ |
| $\ldots$ | $\ldots$ | | $\ldots$ | |
| $n-1$ | $+(n-2)\,I(\xi_{n-1};\eta_0^{n+1}|\xi_0^{n-2})$ | | $\ldots$ | $(\circ)$ |
|   | $+\quad 2\,I(\xi_{n-1};\eta_0^{n+1}|\xi_0^{n-2})$ | $-$ | $2\,I(\xi_{n-1};\eta_0^n|\xi_0^{n-2})$ | $(*)$ |
| $n$ | $+(n-1)\,I(\xi_n\ \ ;\eta_0^{n+1}|\xi_0^{n-1})$ | $-(n-1)\,I(\xi_{n-1};\eta_0^n|\xi_0^{n-2})$ | | $(\circ)$ |
|   | $+\quad I(\xi_n\ \ ;\eta_0^{n+1}|\xi_0^{n-1})$ | $-$ | $I(\xi_n\ \ ;\eta_0^n|\xi_0^{n-1})$ | $(*)$ |
| $n+1$ | $+\quad n\,I(\xi_{n+1};\eta_0^{n+1}|\xi_0^n)$ | $-$ | $n\,I(\xi_n\ \ ;\eta_0^n|\xi_0^{n-1})$ | $(\circ)$ |

TABLE I: Expansion of the sums in (61).

From $\mu \in \mathcal{P}(E_{s_0})$ it follows that $\{\alpha_k, k \in \mathbb{Z}\}$ is an i.i.d.-sequence of random variables. The channel $\kappa$ is assumed to be stationary so that it is in particular $s_0$-stationary according to (2.7.i). Together with the $s_0$-stationarity of $\mu$ we obtain from Lemma 2.9 the $s_0$-stationarity of the channel input-output probability measure $\mu\kappa$. Therefore, the pair sequence $\{(\alpha_k, \beta_k), k \in \mathbb{Z}\}$ of projections is stationary. Applying Lemma 2.16 yields

$$\frac{1}{s_0}I(\alpha_1;\beta_1) \leq \frac{1}{ns_0}I(\alpha_0^n;\beta_0^n) \qquad (69)$$

for all $n \in \mathbb{N}$.

The probability measure $\mu_0$ on $(A_0, \mathcal{A}_0) = (\{\mathtt{a}_{[1]}, \mathtt{a}_{[2]}, \ldots, \mathtt{a}_{[M]}\}, 2^{A_0})$ from which the product measure $\mu \in \mathcal{P}(E_{s_0})$ is built as specified in (15) has the form

$$\mu_0 = \sum_{i=1}^{M} p_i \delta_{\mathtt{a}_{[i]}}. \qquad (70)$$

Then for fixed $n \in \mathbb{N}$ the product measure

$$\mu_0' = \bigotimes_{k=0}^{n-1} \langle\mu_0\rangle_{ks_0} = \sum_{j=1}^{M^n} p_j' \delta_{\mathtt{a}_{[j]}'},$$

with $p_j' = p_{i_0} \cdot p_{i_1} \cdot \ldots \cdot p_{i_{n-1}}$ and $\mathtt{a}_{[j]}' = (\mathtt{a}_{[i_0]}, \mathtt{a}_{[i_1]}, \ldots, \mathtt{a}_{[i_{n-1}]})$ is a probability measure on $(A_0', \mathcal{A}_0')$, where $A_0' = \times_{k=0}^{n-1} \langle A_0\rangle_{ks_0} = \{\mathtt{a}_{[1]}', \ldots, \mathtt{a}_{[M^n]}'\}$ and $\mathcal{A}_0' = \bigotimes_{k=0}^{n-1}\langle\mathcal{A}_0\rangle_{ks_0} = 2^{A_0'}$. We have the identity

$$(A, \mathcal{A}, \mu) = \Big(\underset{l\in\mathbb{Z}}{\times}\langle A_0'\rangle_{lns_0},\ \bigotimes_{l\in\mathbb{Z}}\langle\mathcal{A}_0'\rangle_{lns_0},\ \bigotimes_{l\in\mathbb{Z}}\langle\mu_0'\rangle_{lns_0}\Big)$$

and from $\mathtt{a}_{[j]}' \in \times_{k=0}^{n-1}\langle E_{s_0}\rangle_{ks_0}$ together with the assumption that the family $\mathcal{E}$ of input constraints satisfies the regularity condition (R) in (8) we obtain $\mu \in \mathcal{P}(E_{ns_0})$. This implies for all $n \in \mathbb{N}$

$$\frac{1}{ns_0}I(\alpha_0^n;\beta_0^n) \leq \frac{1}{ns_0}C_{ns_0} \qquad (71)$$

and in combination with (67) and (69) we obtain

$$\rho \leq \limsup_{s\to\infty}\frac{1}{s}C_s \leq \gamma,$$

where the second inequality is obvious from the definition of the limit superior. This proves (23) since $\rho$ was chosen arbitrarily if $\gamma = \infty$ and since we defined $\rho = \gamma - \epsilon$ with $\epsilon > 0$ being arbitrary if $\gamma < \infty$.

(ii) *Proof of (24).* Let us define $\gamma^* = \sup_{\mu\in\mathcal{P}(\mathcal{E})} \bar{I}(\mu)$ and let $\epsilon > 0$. If $\gamma^* < \infty$ we put $\rho^* = \gamma^* - \epsilon$ and if $\gamma^* = \infty$ let $\rho^* > 0$ be arbitrary. Due to the definition of $\gamma^*$ there exists in both cases an $s_0 \in T_+$ and a $\mu \in \mathcal{P}(E_{s_0}) \subset \mathcal{P}(\mathcal{E})$ such that

$$\rho^* \leq \bar{I}(\mu) \leq C. \qquad (72)$$

Indeed, using the notation of part (i) we have

$$\bar{I}(\mu) = \lim_{n\to\infty}\frac{1}{ns_0}I(\xi_0^{ns_0};\eta_0^{ns_0}) \\
= \lim_{n\to\infty}\frac{1}{ns_0}I(\alpha_0^n;\beta_0^n) \quad = \frac{1}{s_0}\bar{I}(\alpha;\beta), \qquad (73)$$

where the existence of the information rate $\bar{I}(\alpha;\beta)$ follows from Lemma 2.16, which can be applied due to the properties of the random sequences $\alpha$ and $\beta$ derived in part (i). Further, using (71) and the representation $C = \sup_{s\in T_+} C_s/s$ yields

$$\frac{1}{ns_0}I(\alpha_0^n;\beta_0^n) \leq \frac{1}{ns_0}C_{ns_0} \leq C \qquad (74)$$

and together with (73) the second inequality in (72). In com-bination with the definition of $\rho^*$ this implies the inequality

$$\gamma^* \leq C. \tag{75}$$

Now consider the setup at the beginning of part (i). Then we have due to (67), (69), and (73)

$$\rho \leq \bar{I}(\mu) \leq \gamma^*$$

and from the definition of $\rho$ follows

$$C \leq \gamma^*. \tag{76}$$

Finally, combining (75) and (76) yields (24).

APPENDIX B. PROOF OF THEOREM 4.7

To prove Theorem 4.7 we need the following properties of the $\psi$-dependence coefficient.

**(B.1) Properties of $\psi$-dependence coefficient.** Let $\alpha$ and $\beta$ be random variables on the probability space $(\Omega, \mathcal{F}, \mathrm{P})$.
  (i) *Nonnegativity:* We have

$$0 \leq \psi(\alpha; \beta),$$

where equality holds if and only if $\alpha$ and $\beta$ are independent.
 (ii) *Monotonicity:* If $\alpha = (\alpha_1, \alpha_2)$ is a random vector, then we have

$$\psi(\alpha_1; \beta) \leq \psi(\alpha; \beta).$$

(iii) $\psi$-*dependence coefficient and Gaussian distribution:* If $(\alpha, \beta)$ is a 2-dimensional Gaussian random vector, then we have

$$\psi(\alpha; \beta) < 2 \implies \mathrm{cor}(\alpha, \beta) = 0,$$

where $\mathrm{cor}(\alpha, \beta)$ denotes the correlation coefficient of $\alpha$ and $\beta$.

Properties (B.1.i) and (B.1.ii) follow directly from the definition of the $\psi$-dependence coefficient. The implication in (B.1.iii) is obtained by combining Prop. 3.4 (d), Prop. 3.11 (a), and Th. 9.7 (I) of [67]. The underlying result for the proof is given in [74, Th. 17.3.2].

**(B.2) Proof of Theorem 4.7.** Using (B.1.i) and (B.1.ii) we obtain with the comment at the beginning of Remark 4.4 the following equivalent definition of finite output memory: The channel $\kappa$ has finite output memory if for all $s \in T$ there exists a $t_o(s) \in T_0$ such that for all $t \geq t_o(s)$ we have

$$\sup_{x \in X} \psi\big(\eta_-^s; \eta_{s+t}^+ \,\big|\, x\big) = 0.$$

Thus, if $\kappa$ has finite output memory, then it is $\psi$-mixing.

Conversely, if $\kappa$ is $\psi$-mixing, then according to Definition 4.3 there exists for any $s \in T$ a $t_o(s) \in T_0$, such that for all $\tilde{t} \geq t_o(s)$ we have

$$\sup_{x \in X} \psi\big(\eta_-^s; \eta_{s+\tilde{t}}^+ \,\big|\, x\big) < 2. \tag{77}$$

Since the family $\{\eta_t, t \in T\}$ of projections is a real $n$-dimensional Gaussian vector process we have $Y_t = Y_{t,1} \times \ldots \times Y_{t,n}$ with $Y_{t,i} = \mathbb{R}$. If $\eta_{t,i}$ denotes the projection from $Y$ onto $Y_{t,i}$, then $\eta_{t,i}$ is a scalar Gaussian random variable and due to (77) and (B.1.ii) we have for all $x \in X$

$$\psi\big(\eta_{t_1,i}; \eta_{t_2,j} \,\big|\, x\big) < 2,$$

which implies together with (B.1.iii)

$$\mathrm{cor}\big(\eta_{t_1,i}, \eta_{t_2,j} \,\big|\, x\big) = 0$$

for all $i, j \in \{1, 2, \ldots, n\}$, $t_1 \leq s$, and $t_2 > s + t_o(s)$. From [67, A901], which is based on the finite-dimensional result in [75, Thm. 9.5.14], we obtain that $\eta_-^s$ and $\eta_{s+t_o(s)}^+$ are independent as random variables on the probability space $(Y, \mathcal{Y}, \kappa(x, \cdot))$ for all $x \in X$. The only-if-part then follows from the equivalent representation of finite output memory given at the beginning of this proof.

APPENDIX C. COUNTEREXAMPLE TO MONOTONICITY PROOF IN [3, APPENDIX II]

Let us adopt the notation and the conditions of Lemma 2.16 and for simplicity let us assume $I(\xi_0^n; \xi_0^n) < \infty$ for all $n \in \mathbb{N}$ in this section. The sequence $\{I(\xi_0^n; \eta_0^n), n \in \mathbb{N}\}$ is nonnegative and monotonically increasing and due to its superadditivity we have $I(\xi_1; \eta_1) \leq I(\xi_0^2; \eta_0^2) - I(\xi_1; \eta_1)$. Therefore, the convexity of the sequence $\{I(\xi_0^n; \eta_0^n), n \in \mathbb{N}\}$ would imply that the sequence $\{n^{-1} I(\xi_0^n; \eta_0^n), n \in \mathbb{N}\}$ is monotonically increasing. This is the argument used in [3, Appendix II]. That it does not work is shown below, where we give a strictly concave example under the conditions of Lemma 2.16.

Note, that the sequence $\{I(\xi_0^n; \eta_0^n), n \in \mathbb{N}\}$ is convex if

$$I(\xi_0^n; \eta_0^n) \leq \frac{1}{2}\Big[I(\xi_0^{n-1}; \eta_0^{n-1}) + I(\xi_0^{n+1}; \eta_0^{n+1})\Big] \tag{78}$$

or equivalently if

$$0 \leq \Big[I(\xi_0^{n+1}; \eta_0^{n+1}) - I(\xi_0^n; \eta_0^n)\Big] \\ - \Big[I(\xi_0^n; \eta_0^n) - I(\xi_0^{n-1}; \eta_0^{n-1})\Big] \tag{79}$$

holds for all $n = 2, 3, \ldots$. It is strictly convex if the inequality in (78) or (79) is sharpened to a strict inequality. Furthermore, the sequence $\{I(\xi_0^n; \eta_0^n), n \in \mathbb{N}\}$ is (strictly) concave if the sequence $\{-I(\xi_0^n; \eta_0^n), n \in \mathbb{N}\}$ is (strictly) convex.

To evaluate the difference on the right-hand side of (79) we rewrite $I(\xi_0^n; \eta_0^n)$ as

$$\begin{aligned} I(\xi_0^n; \eta_0^n) &= I(\xi_1; \eta_1) + I(\xi_1^n; \eta_1^n) && + I(\xi_1^n; \xi_1, \eta_1 \,|\, \eta_1^n) \\ & \quad - I(\xi_1; \xi_1^n) && + I(\xi_1; \eta_1^n \,|\, \eta_1) \\ &= I(\xi_1; \eta_1) + I(\xi_0^{n-1}; \eta_0^{n-1}) + I(\xi_1, \eta_1; \xi_1^n, \eta_1^n) \\ & \qquad\qquad\qquad\qquad\qquad\qquad - I(\eta_1; \eta_1^n). \end{aligned}$$

The first equality follows from applying the chain rules of (conditional) mutual information multiple times. For the second equality we use the stationarity of the pair sequence $\{(\xi_k, \eta_k), k \in \mathbb{Z}\}$, the independence of $\xi_1$ and $\xi_1^n$ implying $I(\xi_1; \xi_1^n) = 0$, and again the chain rule. Using this representa-

tion, we obtain

$$\left[I(\xi_0^{n+1};\eta_0^{n+1}) - I(\xi_0^n;\eta_0^n)\right] - \left[I(\xi_0^n;\eta_0^n) - I(\xi_0^{n-1};\eta_0^{n-1})\right]$$
$$= \left(I(\xi_1,\eta_1;\xi_1^{n+1},\eta_1^{n+1}) - I(\xi_1,\eta_1;\xi_1^n,\eta_1^n)\right)$$
$$\quad - \left(I(\eta_1;\eta_1^{n+1}) - I(\eta_1;\eta_1^n)\right)$$
$$= I(\xi_1,\eta_1;\xi_{n+1},\eta_{n+1}|\xi_1^n,\eta_1^n) - I(\eta_1;\eta_{n+1}|\eta_1^n), \quad (80)$$

where the last equality follows from the chain rule. Thus, we can use the difference in (80) to analyze the convexity or concavity of the sequence $\{I(\xi_0^n;\eta_0^n), n \in \mathbb{N}\}$. Below we give an example where the difference is negative for all $n = 2, 3, \ldots$, i.e., where the sequence $\{I(\xi_0^n;\eta_0^n), n \in \mathbb{N}\}$ is strictly concave.

Let $\{\xi_k, k \in \mathbb{Z}\}$ be an i.i.d.-sequence, where $\xi_0$ has a Bernoulli distribution with the parameter $q = \mathrm{P}(\xi_0 = 1)$. Assume that the random sequence $\{\eta_k, k \in \mathbb{Z}\}$ is defined by the relation

$$\eta_k = \xi_k \oplus \xi_{k-1} \quad (81)$$

for all $k \in \mathbb{Z}$, where $\oplus$ denotes addition modulo 2. The stationarity of the pair sequence $\{(\xi_k, \eta_k), k \in \mathbb{Z}\}$ is an immediate consequence of this construction. Thus the conditions of Lemma 2.16 are satisfied.

It is not difficult to see that we have the Markov chain $\bigl((\xi_1, \eta_1) - (\xi_1^n, \eta_1^n) - (\xi_{n+1}, \eta_{n+1})\bigr)$, which implies that the first term in (80) is zero such that

$$I(\xi_1, \eta_1; \xi_{n+1}, \eta_{n+1} | \xi_1^n, \eta_1^n) - I(\eta_1; \eta_{n+1} | \eta_1^n)$$
$$= -I(\eta_1; \eta_{n+1} | \eta_1^n) \le 0$$

holds for all $m = 2, 3, \ldots$, since the conditional mutual information is nonnegative. Thus, the sequence $\{I(\xi_0^n;\eta_0^n), n \in \mathbb{N}\}$ is concave. It is strictly concave if $I(\eta_1; \eta_{n+1}|\eta_1^n) > 0$, i.e., if the Markov chain relation $(\eta_1 - \eta_1^n - \eta_{n+1})$ does not hold. That this is the case for parameter $q \in (0, 1/2) \cup (1/2, 1)$ is most easily obtained considering the conditional probabilities for the sequence of all zeros. This example shows the insufficiency of the convexity argument to prove the monotonicity of the sequence $\{n^{-1}I(\xi_0^n;\eta_0^n), n \in \mathbb{N}\}$. The channel representation of this example, which is stationary and memoryless (see Remark 4.4) is discussed in [5, Par. 16.6].

ACKNOWLEDGMENT

The authors wish to thank Holger Boche for helpful discussions and valuable comments, in particular regarding the optimistic and pessimistic view on coding capacities as considered in Paragraph 2.13 and Remark 3.3. Furthermore, the authors would like to thank the anonymous reviewers for their technical and editorial comments, which significantly helped to improve the presentation of the results.

**Martin Mittelbach** was born in Leinefelde, Germany, in 1978. In May 2003 he received a diploma degree in Electrical Engineering with a major in Information Technologies and in December 2012 a diploma degree in Mathematics with a major in Mathematical Stochastics, both from Dresden University of Technology, Germany. In December 2014 he completed his dissertation at the Chair of Communications Theory, Dresden University of Technology and received the academic degree Doktor-Ingenieur. Since then he holds a post-doctoral position there. Martin's main research interests are within the fields of applied probability theory and information theory.

**Eduard A. Jorswieck** (S'01–M'05–SM'08) was born in 1975 in Berlin, Germany. He received his Ph.D. degree in electrical engineering and computer science from the Technische Universität Berlin, Germany, in 2004. Since February 2008, he has been the head of the Chair of Communications Theory and Full Professor at Dresden University of Technology (TUD), Germany. Eduard's main research interests are in the area of signal processing for communications and networks, applied information theory, and communications theory. He has published more than 100 journal papers, 11 book chapters, 3 monographs, and some 250 conference papers on these topics. Dr. Jorswieck is senior member of IEEE. Since 2017 he acts as Editor-in-Chief for the Springer EURASIP Journal on Wireless Communications and Networking. Since 2015, he is member of the IEEE SAM TC. Since 2016, he serves as Associate Editor for IEEE Transactions on Information Forensics and Security. He has served on the editorial boards of IEEE Signal Processing Letters, IEEE Transactions on Signal Processing and IEEE Transactions on Wireless Communications. In 2006, he received the IEEE Signal Processing Society Best Paper Award.